\PassOptionsToPackage{noend}{algpseudocode}
\documentclass[unnumsec,webpdf,contemporary,large,numbered]{oup-authoring-template}

\usepackage{amsmath,amssymb,booktabs,graphicx,flafter}
\usepackage{algorithm}
\usepackage{algpseudocode}

\makeatletter
\let\ps@headings\ps@plain
\let\ps@opening\ps@plain
\makeatother
\begin{document}

\firstpage{1}

\title[MAPS for EHR Survival]{Multivariate Augmented Prediction-Powered Survival Inference for EHR Association Studies With Limited Chart Review}

\author[1]{Xicheng Xie}
\author[2]{Katherine Fiona Simpson}
\author[1]{Hyojung Jang}
\author[2,$\dagger$]{Theresa L. Walunas}
\author[1,$\ast$,$\dagger$]{Lili Zhao}

\address[1]{\orgdiv{Division of Biostatistics, Department of Preventive Medicine}, \orgname{Northwestern University Feinberg School of Medicine}, \orgaddress{Chicago, \state{IL} \postcode{60611}, \country{USA}}}
\address[2]{\orgdiv{Division of General Internal Medicine, Department of Medicine}, \orgname{Northwestern University Feinberg School of Medicine}, \orgaddress{Chicago, \state{IL} \postcode{60611}, \country{USA}}}

\corresp[$\ast$]{Corresponding author: Lili Zhao. \href{mailto:zhaolili@northwestern.edu}{zhaolili@northwestern.edu}}
\corresp{\quad\textsuperscript{$\dagger$}Theresa L. Walunas and Lili Zhao are co-senior authors.}

\abstract{\mdseries
\textbf{Objective:} To reduce bias and improve precision in survival association studies using electronic health record (EHR) data when computational phenotypes (CPs) for outcomes and covariates are error-prone and chart review is limited.
\\[3pt]
\textbf{Materials and Methods:} We developed Multivariate Augmented Prediction-Powered Survival (MAPS) regression, which combines chart-reviewed and CP-derived labels across different validation patterns. MAPS uses a discrete-time survival model with a complementary log-log link to estimate hazard ratios (HRs). It incorporates data-adaptive augmentation weights, inverse-probability sampling weights, and patient-level cluster-robust variance estimation. We compared MAPS with chart-reviewed data (CRD) and CP-only analyses in simulation studies and a masked-validation study of immune-related adverse events (irAEs) among 1,026 patients receiving immune checkpoint inhibitors.
\\[3pt]
\textbf{Results:} In the main simulations, MAPS produced approximately unbiased estimates and generally near-nominal 95\% confidence interval (CI) coverage, with narrower intervals than CRD. CP-only estimates became increasingly biased as CP errors increased. In the EHR analysis, MAPS estimates closely agreed with fully validated estimates and improved precision over CRD, particularly at lower validation rates. CP-only analysis suggested increased irAE risk associated with pre-existing autoimmune disease (AID; HR=1.53; 95\% CI, 1.12--2.09), whereas the fully validated analysis did not (HR=0.93; 95\% CI, 0.67--1.29).
\\[3pt]
\textbf{Discussion:} Precision gains depended on CP quality and validation rates, while small validation samples may affect confidence interval coverage.
\\[3pt]
\textbf{Conclusion:} MAPS provides a flexible and efficient approach to EHR-based survival inference by combining limited chart-reviewed data with error-prone CPs across different validation patterns.
}

\keywords[Abbreviations]{AID, CP, CRD, EHR, HR, ICI, IPW, irAE, MAPS}

\maketitle
\section{Background and Significance}

Electronic health records (EHRs) support longitudinal studies of disease risk and treatment-related adverse events, but many clinical variables must first be extracted as computational phenotypes (CPs). CP algorithms translate diagnosis codes, laboratory results, medications, and clinical narratives into research variables using clinical rules, machine learning, or natural language processing \citep{carrell2024framework}. The resulting phenotypes are imperfect measurements. Treating them as observed truth can bias association estimates, reduce statistical power, and, under differential misclassification, inflate type I error \citep{duan2016empirical,tong2020augmented}. Chart review and clinical adjudication provide validated labels, but their cost often limits validation to a small subset of patients \citep{yin2022cost}. Restricting analyses to this subset reduces statistical power and fails to leverage potentially valuable information from the remaining cohort.

Several methods have been developed to address phenotyping error in EHR-based association studies. Early likelihood-based approaches incorporate known or estimated sensitivity and specificity to correct for outcome misclassification \citep{magder1997logistic}. Hubbard et al. proposed a bias-correction method that rescales regression coefficients using the mean predicted phenotype probabilities among true cases and controls \citep{hubbard2020reducing}. Tong et al. developed an augmented estimator that combines chart-reviewed labels with CPs to reduce bias and improve precision \citep{tong2020augmented}. Subsequent extensions accommodate outcome-dependent chart-review sampling \citep{yin2022cost} and multiple CPs generated by different algorithms for the same outcome \citep{lu2024leveraging}. More recently, prediction-powered inference (PPI) extended the augmentation framework to a broader range of statistical problems, including mean and quantile estimation and generalized linear regression \citep{angelopoulos2023prediction}. PPI++ further introduces data-adaptive weighting to optimize the contribution of predicted outcomes, improving statistical efficiency, and provides computationally efficient inference \citep{angelopoulos2023ppiplusplus}. However, these methods do not directly accommodate censored time-to-event outcomes or simultaneously address phenotyping errors in both covariates and outcomes.

Few methods address measurement error in survival regression. Existing approaches use validation-based multiple imputation or imputation-assisted generalized raking to correct measurement errors and improve efficiency \citep{giganti2020accounting,oh2021improved}. Multiple imputation requires an adequately specified imputation model, whereas imputation-assisted raking can remain consistent under model misspecification. Small validation samples can lead to imputation-model overfitting or limited efficiency gains.

To address these limitations, we propose Multivariate Augmented Prediction-Powered Survival (MAPS) regression, a framework that combines validated and CP-derived labels to estimate covariate associations with survival outcomes when both outcomes and covariates are subject to phenotyping error. MAPS uses discrete-time survival regression with a complementary log-log link to estimate hazard ratios (HRs) \citep{prentice1978grouped}. Built on the recent work on augmented estimation under different missing-data patterns \citep{chen2025generalmissing}, MAPS partitions patients according to their label-validation patterns and constructs pattern-specific augmentation terms to incorporate information from partially validated and CP-derived labels. A data-adaptive weight matrix optimally combines these augmentation terms to improve statistical efficiency. Inverse-probability sampling weights account for covariate-dependent validation, while patient-level cluster-robust variance estimation accounts for within-patient dependence across follow-up intervals.

We evaluate MAPS through simulation studies with CP errors in the survival outcome alone and in both the outcome and a covariate. We assess bias, empirical coverage of 95\% confidence intervals, and precision across varying levels of CP error, comparing MAPS regression with chart-reviewed data (CRD) and CP-only analyses. We further evaluate MAPS using a Northwestern Medicine cohort of patients receiving immune checkpoint inhibitors (ICIs), examining the association between pre-existing autoimmune disease (AID) and the risk of immune-related adverse events (irAEs). Because both AID history and irAE outcomes have been fully adjudicated in this cohort, we mask validated labels to vary the validation rate and assess how closely MAPS reproduces estimates from the fully validated cohort and the precision gains achieved across different validation rates.

\section{Materials and Methods}

\subsection{Discrete-Time Survival Regression Models}
Consider a study with $N$ subjects. For subject $i$, let $X_i$ denote a vector of covariates, $T_i$ the event time, and $C_i$ the censoring time. In practice, neither $T_i$ nor $C_i$ is directly observed. Instead, we observe the follow-up time $U_i=\min(T_i,C_i)$ and the event indicator $\Delta_i=I(T_i\le C_i)$.

A widely used model for assessing the association between covariates and time-to-event outcomes is the Cox proportional hazards model,
\begin{equation}
\lambda(t\mid X_i)=\lambda_0(t)\exp(\beta^\top X_i),
\label{eq:cox-ph}
\end{equation}
where $\lambda_0(t)$ is the baseline hazard function and $\beta$ is a vector of regression coefficients that quantify the association between the covariates and the hazard of the event.

In a discrete-time survival model, follow-up is divided into prespecified intervals, such as months or years. Let $J_i$ denote the number of intervals contributed by subject $i$. For $j=1,\ldots,J_i$, define subject $i$'s event indicator as
\begin{equation}
d_{ij}=
\begin{cases}
1, & \text{event during interval }j,\\
0, & \text{otherwise}.
\end{cases}
\label{eq:interval-event}
\end{equation}
A subject who is censored during time interval $j$ has $d_{ij}=0$ for that interval and does not contribute to subsequent intervals. We define the discrete-time hazard for subject $i$ during interval $j$ as
\begin{equation}
h_{ij}= \Pr\left(d_{ij}=1\mid \text{at risk at the start of interval }j,X_i\right).
\label{eq:discrete-hazard}
\end{equation}
Thus, $h_{ij}$ is the conditional probability that subject $i$ experiences the event during interval $j$, given that the subject has remained event-free and uncensored up to the start of that interval.

The data are represented in person-period format, with one row for each time interval during which a subject remains at-risk for the event. We fit a regression model with a complementary log-log (cloglog) link to these data. Let $b_j\in\mathbb R^q$ denote a flexible basis function for the intervals and let $X_i\in\mathbb R^r$ denote the covariates. We specify
\begin{equation}
\log\{-\log(1-h_{ij})\}=b_j^\top\alpha+X_i^\top\beta,
\label{eq:cloglog}
\end{equation}
where $\alpha$ indicates the effect of baseline time. With the cloglog link, $\beta$ contains the log hazard ratios (log-HRs) for the covariates. With sufficiently small time intervals, the estimated hazard ratios closely approximate those obtained from continuous-time Cox regression \citep{prentice1978grouped} (Supplementary Methods, Section~S1).

\subsection{Multivariate Augmented Prediction-Powered Survival (MAPS) Regression}

\begin{figure*}[t]
\centering
\includegraphics[width=\textwidth]{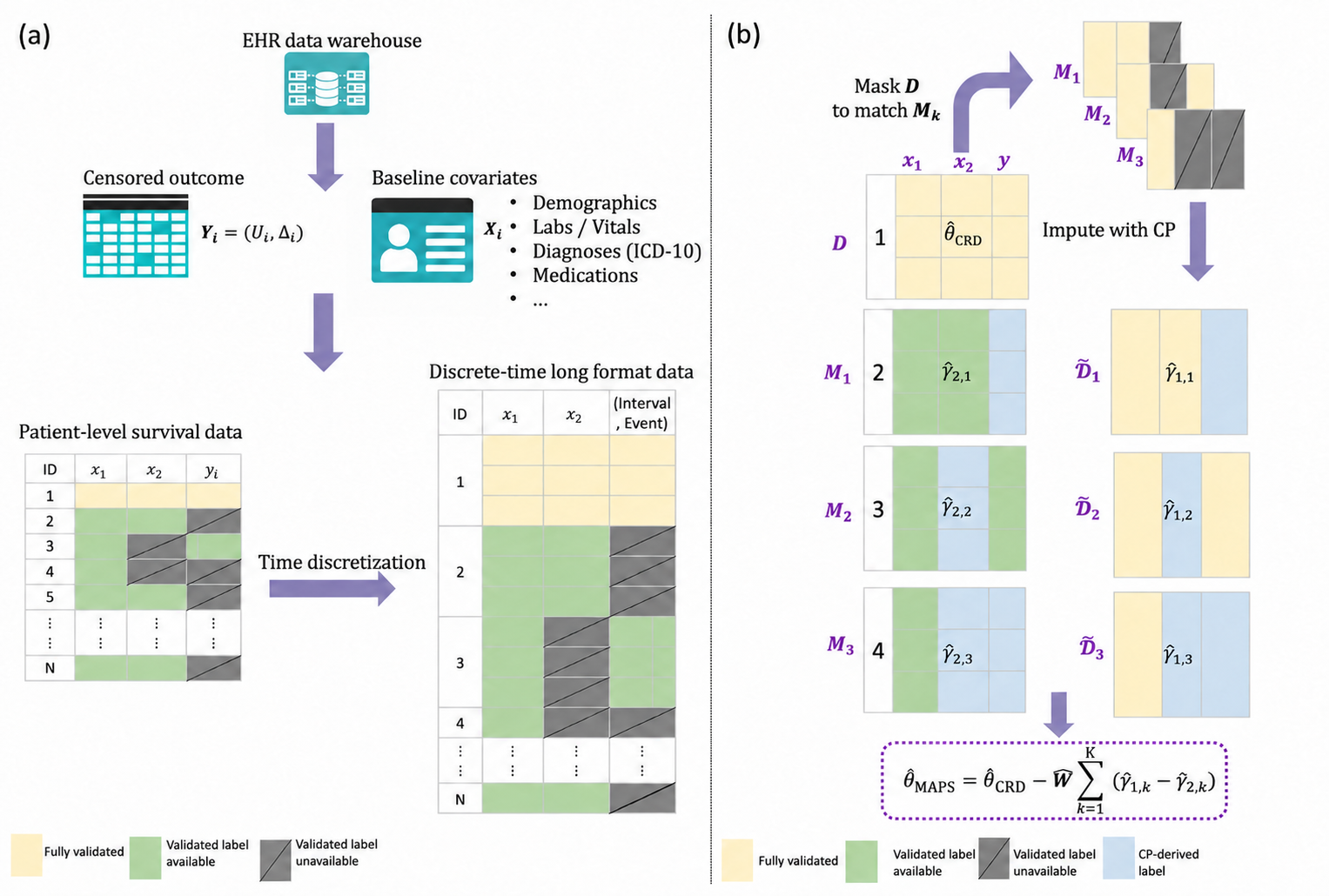}
\caption{Overview of MAPS for discrete-time survival analysis. (a) Patient-level survival data are expanded into one record per observed follow-up interval. (b) The cohort is stratified into the fully validated dataset $\mathcal D\equiv\mathcal M_0$ and datasets $\mathcal M_k$, $k=1,2,3$, in which CPs replace unavailable validated labels. Corresponding labels in $\mathcal D$ are masked and replaced with their CP-derived counterparts to form $\widetilde{\mathcal D}_k$. Differences between estimates from $\widetilde{\mathcal D}_k$ and $\mathcal M_k$ form pattern-specific rectifiers that augment the CRD estimate through a data-adaptive weight matrix.}
\label{fig:debias}
\end{figure*}

We use the term \emph{validated label} to denote a covariate or outcome label determined through chart review and clinical adjudication and treated as the true label for analysis. We first consider the setting illustrated in Figure~\ref{fig:debias}, in which the survival outcome $Y_i=(U_i,\Delta_i)$ and one covariate, $X_{i1}$, are identified using CP algorithms. These CP-derived labels, denoted by $\widehat{Y}_i$ and $\widehat{X}_{i1}$, are available for all subjects but are measured with error. We assume that the remaining covariates, $X_{i,-1}$, are measured without error for all subjects.

Our goal is to estimate the $p$-dimensional parameter $\theta=(\alpha^\top,\beta^\top)^\top$, where $p=q+r$, $\alpha$ contains regression coefficients for baseline time, and $\beta$ contains log-HRs for the covariates. MAPS uses a subset of subjects with validated labels to correct the bias that can arise from directly using error-prone CP-derived outcomes and covariates, while incorporating CP-derived labels from the full cohort to improve efficiency. The key idea of MAPS regression is to partition the full cohort into mutually exclusive subsets defined by distinct label validation patterns and construct a debiased estimator by combining information across these subsets.

\textbf{Dataset preparation for debiased estimation.}
Based on label validation status, we partition the overall cohort into four mutually exclusive datasets:
\begin{itemize}
    \item $\mathcal M_0\equiv\mathcal D$ contains validated labels for both $X_1$ and $Y$;
    \item $\mathcal M_1$ contains validated labels for $X_1$ and the CP-derived label $\widehat{Y}$ for $Y$;
    \item $\mathcal M_2$ contains validated labels for $Y$ and the CP-derived label $\widehat{X}_1$ for $X_1$; 
    \item $\mathcal M_3$ contains CP-derived labels $\widehat{X}_1$ and $\widehat{Y}$ for $X_1$ and $Y$, respectively.
\end{itemize}
Each dataset $\mathcal M_k$ represents a distinct validation pattern. If only the outcome is subject to CP error, there are two datasets with distinct validation patterns, $\mathcal M_0$ and $\mathcal M_1$. More generally, if $s$ variables are subject to CP error, there are up to $2^s$ distinct validation-pattern datasets.

For each dataset containing CP-derived labels ($\mathcal M_k$, $k=1,2,3$), we construct a corresponding \emph{mask-and-impute} dataset $\widetilde{\mathcal D}_k$ from the fully validated dataset $\mathcal D$. Specifically, we mask the validated labels that are unavailable in $\mathcal M_k$ and replace them with their CP-derived counterparts, thereby reproducing the same validation pattern. Thus, $\widetilde{\mathcal D}_1$ replaces $Y$ with $\widehat{Y}$, $\widetilde{\mathcal D}_2$ replaces $X_1$ with $\widehat{X}_1$, and $\widetilde{\mathcal D}_3$ replaces both $X_1$ and $Y$ with $\widehat{X}_1$ and $\widehat{Y}$, respectively.

\textbf{De-biased estimation for survival regression.}
MAPS estimates the regression parameters through the following estimating equations:
\begin{align}
\sum_{i\in\mathcal D}
a_i^{(0)}S_i(\widehat\theta_{\mathrm{CRD}};\mathcal D_i)&=0,
\label{eq:wcc-ee}\\
\sum_{i\in\mathcal D}
a_i^{(0)}S_i(\widehat\gamma_{1,k};\widetilde{\mathcal D}_{k,i})&=0,
\qquad k=1,2,3,
\label{eq:mask-impute-ee}\\
\sum_{i\in\mathcal M_k}
a_i^{(k)}S_i(\widehat\gamma_{2,k};\mathcal M_{k,i})&=0,
\qquad k=1,2,3.
\label{eq:pattern-ee}
\end{align}
Here, $S_i(\vartheta;\mathcal E_i)$ denotes the $p$-dimensional subject-level estimating function for the discrete-time survival model with a cloglog link, obtained by summing contributions across all observed at-risk intervals for subject $i$. $a_i^{(k)}$ is the sampling weight for subject $i$ in dataset $\mathcal M_k$. The CRD estimator $\widehat\theta_{\mathrm{CRD}}$ is obtained from the fully validated dataset $\mathcal D$. For each validation pattern $k=1,2,3$, the auxiliary estimators $\widehat\gamma_{1,k}$ and $\widehat\gamma_{2,k}$ are obtained from the corresponding mask-and-impute dataset $\widetilde{\mathcal D}_k$ and validation-pattern dataset $\mathcal M_k$, respectively. All parameter estimates are $p$-dimensional.

If chart review is conducted using simple random sampling, we set $a_i^{(k)}=1$ for all subjects across all datasets. If selection for chart review depends on observed patient characteristics, $a_i^{(k)}$ is estimated as the inverse probability of belonging to validation-pattern dataset $\mathcal M_k$:
\begin{equation}
a_i^{(k)}
=
\frac{1}{\Pr\!\left(i\in\mathcal M_k\mid Z_i\right)},
\qquad k=0,1,2,3,
\label{eq:ipw}
\end{equation}
where $Z_i$ denotes the observed variables associated with the chart review selection process. Stabilized or truncated weights may be used to reduce the influence of extreme weights \citep{sun2018inverse}.

The de-biased estimator in MAPS regression is:
\begin{equation}
\begin{aligned}
\widehat\theta_{\mathrm{MAPS}}
&=
\widehat\theta_{\mathrm{CRD}}
-\widehat W
\sum_{k=1}^{3}
\left(
\widehat\gamma_{1,k}-\widehat\gamma_{2,k}
\right).
\end{aligned}
\label{eq:maps-estimator}
\end{equation}
Here, $\widehat\Delta_k=\widehat\gamma_{1,k}-\widehat\gamma_{2,k}$ is the rectifier associated with validation-pattern dataset $\mathcal M_k$. The weighted sum of these rectifiers is the augmentation term in Equation~\eqref{eq:maps-estimator}, and $\widehat W$ is a $p\times p$ weight matrix estimated by
\begin{equation}
\widehat W
=
\widehat\Sigma_{\theta,\Delta}
\widehat\Sigma_{\Delta}^{-1},
\label{eq:optimal-weight}
\end{equation}
where $\widehat\Sigma_{\theta,\Delta}=\widehat{\operatorname{Cov}}(\widehat\theta_{\mathrm{CRD}},\widehat\Delta)$ and $\widehat\Sigma_{\Delta}=\widehat{\operatorname{Var}}(\widehat\Delta),$ with $\widehat\Delta=\sum_{k=1}^{3}\widehat\Delta_k$.

The covariance of $\widehat\theta_{\mathrm{MAPS}}$ is estimated as
\begin{equation}
\widehat\Sigma_{\mathrm{MAPS}}
=
\widehat\Sigma_{\theta}
-
\widehat W
\widehat\Sigma_{\Delta,\theta}.
\label{eq:maps-variance}
\end{equation}
Here, $\widehat\Sigma_\theta\in\mathbb R^{p\times p}$ is the estimated covariance of $\widehat\theta_{\mathrm{CRD}}$ and $\widehat\Sigma_{\Delta,\theta}=\widehat\Sigma_{\theta,\Delta}^\top$. The matrix $\widehat W$ is chosen to minimize the asymptotic variance of the MAPS estimator assuming a common weight matrix for all three datasets. Under the stated assumptions, the MAPS estimator has asymptotic variance no greater than that of $\widehat\theta_{\mathrm{CRD}}$ \citep{chen2025generalmissing}.

The paired construction aligns the inferential targets of
$\widetilde{\mathcal D}_k$ and $\mathcal M_k$ by using the same
combination of validated and CP-derived variables. Under correct
sampling weights and an ignorable validation mechanism,
$\widehat\gamma_{1,k}$ and $\widehat\gamma_{2,k}$ converge to
the same population parameter, implying
$\widehat\Delta_k\xrightarrow{p}0$.
Thus, incorporating the augmentation terms preserves the
consistency of $\widehat\theta_{\mathrm{CRD}}$ while potentially
improving efficiency by leveraging CP-derived information
from the larger cohort.

For intuition, consider a single validation pattern with a scalar augmentation weight $W$. If the CP-derived labels provide no useful information for estimating the target parameter, such that $\Sigma_{\theta,\Delta}=0$, the optimal weight is $W=0$, and MAPS reduces to $\widehat\theta_{\mathrm{CRD}}$, using only validated data. As the CP-derived labels become more informative, the augmentation can yield greater variance reduction. At the other extreme, if the CP-derived labels perfectly match the true labels, MAPS is asymptotically equivalent to the full-cohort estimator under random validation and comparable information per subject.

To account for repeated time intervals within subjects in the discrete-time survival model, we use a cluster-robust sandwich variance estimator, aggregating interval-specific estimating functions within each patient. This accounts for within-patient correlation across follow-up intervals and correlations among estimates obtained from the same fully validated subset $\mathcal D$ (Supplementary Methods, Section~S7). We construct 95\% confidence intervals (CIs) on the log-HR scale using standard errors obtained from the square roots of the diagonal elements of the estimated covariance matrix, and then exponentiate the confidence limits to obtain CIs for HRs.

\begin{algorithm}[H]
\caption{Procedure for estimating MAPS regression.}
\label{alg:maps}
\begin{algorithmic}[1]
\Require CP-derived covariates and survival outcomes for all subjects and validated covariates and outcomes for a subset of subjects.
\State Divide the full cohort into $\mathcal D,\mathcal M_1,\mathcal M_2,\mathcal M_3$ based on label-validation pattern (assuming CP errors in one covariate and outcome).

\State Estimate $\widehat\theta_{\mathrm{CRD}}$ from the fully validated dataset $\mathcal D$.
\State For each $k=1,2,3$, construct $\widetilde{\mathcal D}_k$ to match $\mathcal M_k$, estimate $\widehat\gamma_{1,k}$ from $\widetilde{\mathcal D}_k$, and estimate $\widehat\gamma_{2,k}$ from $\mathcal M_k$, using sampling weights if needed.
\State Estimate $\widehat W$ and $\widehat\theta_{\mathrm{MAPS}}=\widehat\theta_{\mathrm{CRD}}-\widehat W\sum_{k=1}^3(\widehat\gamma_{1,k}-\widehat\gamma_{2,k})$.
\State Report the estimated baseline hazard function and covariate hazard ratios with 95\% confidence intervals.
\end{algorithmic}
\end{algorithm}

\section{Simulation Design and Evaluation}

We conducted simulation studies to evaluate the performance of MAPS under two settings: (1) CP error in the survival outcome alone, and (2) CP error in both the survival outcome and a covariate. In each setting, we generated a cohort of $N=5000$ subjects with covariates $X_i=(x_{i1},x_{i2})^\top$ and event times from a Weibull proportional hazards model:
\begin{equation}
T_i\mid x_{i1},x_{i2}
\sim\operatorname{Weibull}\!\left(
\kappa,\;
\lambda\exp\!\left\{
-\frac{\beta_1x_{i1}+\beta_2x_{i2}}{\kappa}
\right\}
\right).
\label{eq:sim-weibull}
\end{equation}
We set the shape to $\kappa=1.5$ and baseline scale parameter to $\lambda=3$. The coefficients $\beta_1=\log(1.4)$ and $\beta_2=\log(0.75)$ corresponded to HRs of 1.4 and 0.75, respectively. We generated exponential censoring times independently of event times conditional on the covariates, with the rate calibrated to yield 35\% censoring. The observed follow-up times $U_i=\min(T_i,C_i)$ ranged from near 0 to approximately 21 time units across simulated cohorts.

CP-derived event times were generated from a perturbed Weibull regression model by adding an error term to the linear predictor:
\begin{equation}
\widetilde\ell_i=\beta_1x_{i1}+\beta_2x_{i2}+\varepsilon_i,
\qquad
\varepsilon_i\overset{\mathrm{ind}}{\sim}N(\delta,\sigma^2),
\label{eq:sim-cp-linear-predictor}
\end{equation}
where $\delta$ controls the systematic shift and $\sigma$ controls the random variation in the linear predictor. The errors were generated independently of the covariates. True and CP-derived event times shared the same subject-specific latent Weibull draw and censoring time.

\textbf{Generating CP errors.} In simulations with CP errors only in the outcomes, covariates $(x_1,x_2)$ were generated from a bivariate normal distribution with zero means, unit marginal variances, and correlation 0.2. Both covariates were measured without error for all subjects. CP-derived survival outcomes were available for the full cohort, while validated outcomes were available for approximately 10\% of subjects. We fixed $\delta=0.30$ and varied $\sigma$ to evaluate performance across different levels of random prediction error.

In simulations with CP errors in both covariates and outcomes, we generated $x_1\sim N(0,1)$ and $x_2\sim\operatorname{Bernoulli}\{\operatorname{expit}(0.6x_1)\}$. The covariate $x_1$ was measured without error, whereas $x_2$ and the survival outcome were subject to CP errors. CP-derived labels for $x_2$ were generated by independently flipping its binary value with a prespecified probability. We varied this misclassification probability from 0\% to 30\%, with outcome error parameters fixed at $\delta=0$ and $\sigma=0.05$. Validation rates were 40\% for $x_2$ and 25\% for the survival outcome.

\textbf{Subject selection for chart review.} We used logistic regression to assign subjects to different validation label patterns. For error in the outcome only, we used
\begin{equation}
\operatorname{logit}\Pr(R_{Yi}=1\mid X_i)
=c_0+0.6x_{i1}-0.6x_{i2}.
\label{eq:sim-ps-outcome}
\end{equation}
For error in both a covariate and the outcome, we specified
\begin{equation}
\operatorname{logit}\Pr(R_{Xi}=1\mid x_{i1})
=a_0+0.6x_{i1},
\label{eq:sim-ps-covariate}
\end{equation}
\begin{equation}
\begin{aligned}
&\operatorname{logit}\Pr(R_{Yi}=1\mid x_{i1},R_{Xi})\\
&\quad=b_0+0.6x_{i1}-0.8(1-R_{Xi}).
\end{aligned}
\label{eq:sim-ps-outcome-multivariate}
\end{equation}

Here, $R_{Yi}=1$ indicates that subject $i$ was selected for outcome validation, and $R_{Xi}=1$ indicates selection for validation of covariate $x_2$. We drew these indicators from Bernoulli distributions with probabilities specified by the logistic models, producing two validation patterns for outcome error alone and four patterns, $\mathcal M_0$--$\mathcal M_3$, for joint error. We fixed the slope coefficients at the specified values and calibrated the intercepts to achieve the desired validation rates. For analysis, we fitted the same propensity-score models to validated-label availability to estimate validation-pattern probabilities, bounded them below at 0.01, and constructed stabilized inverse-probability sampling weights. We present results under covariate-dependent sampling. Additional analyses of validation patterns and sample size are provided in the Supplementary Methods.

We compared the proposed MAPS regression with two alternative approaches:
(1) chart-reviewed data (CRD) analysis, which included the subset of subjects
with fully validated data for all variables;
and (2) CP-only analysis, which used CP-derived labels without validation.

All analyses used a time interval width of $dt=0.5$ in the discrete-time survival model and a natural spline basis with 3 degrees of freedom to model baseline time. Stabilized inverse-probability sampling weights were applied to MAPS and CRD to account for covariate-dependent sampling; CP-only analysis used unit weights. Each simulation scenario was replicated 300 times. We evaluated performance in terms of bias, empirical coverage of 95\% confidence intervals (CIs), and average CI width for each log-HR coefficient. Bias was defined as the mean estimated coefficient minus its true value, coverage as the proportion of 95\% CIs containing the true value, and CI width as the average width of the 95\% CIs across replicates. Additional simulation settings and implementation details are provided in the Supplementary Methods.

\subsection{Simulation Results}

\begin{figure}[!t]
\centering
\includegraphics[width=\linewidth]{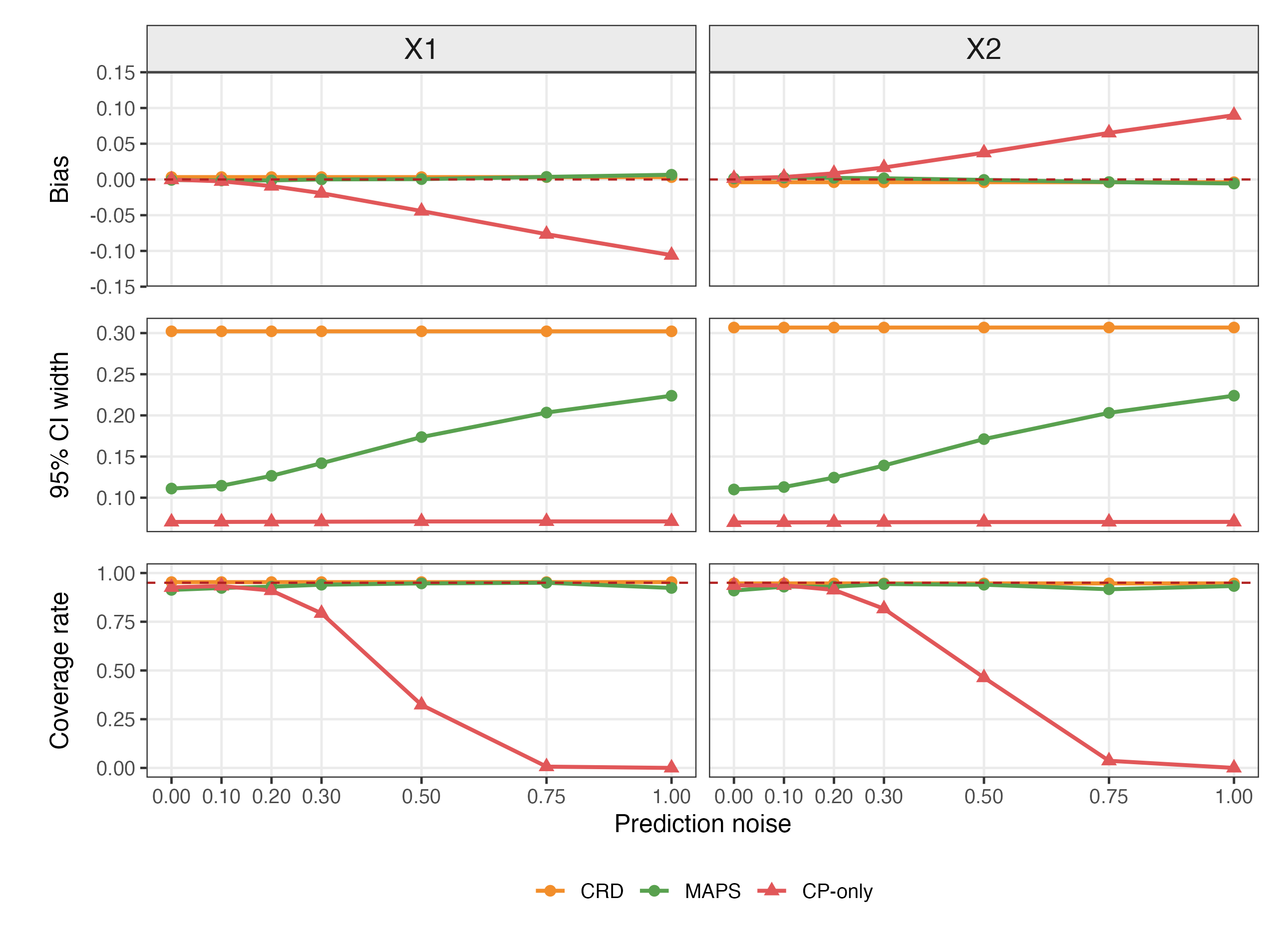}
\caption{Performance across prediction-noise levels in the outcome CP ($\sigma$: 0--1), with $\delta=0.30$, $N=5000$, 10\% outcome validation, and 300 replicates per level. Bias and mean CI width are on the log-HR scale. Dashed red lines indicate zero bias and nominal 95\% coverage. CRD, chart-reviewed data analysis.}
\label{fig:sim_missingoutcome_noise}
\end{figure}

\textbf{CP Error in the Survival Outcome.}
As shown in Figure~\ref{fig:sim_missingoutcome_noise}, MAPS and CRD remained approximately unbiased for both coefficients across all levels of prediction noise. However, MAPS consistently achieved greater precision, with substantially narrower 95\% CIs than CRD. Although MAPS intervals widened as prediction noise increased, they remained narrower than CRD intervals even at the highest noise level. Both methods maintained generally near-nominal 95\% coverage (91\%--95\% for MAPS and approximately 95\% for CRD). In contrast, CP-only estimates exhibited increasing bias as prediction noise increased. Despite their narrow CIs, CP-only estimates failed to account for this bias, resulting in sharply declining coverage probabilities that approached 0\% for both coefficients at the highest noise level. Overall, MAPS substantially improved precision over CRD while maintaining negligible bias and near-nominal coverage, consistently outperforming CP-only estimation.

\begin{figure}[!t]
\centering
\includegraphics[width=\linewidth]{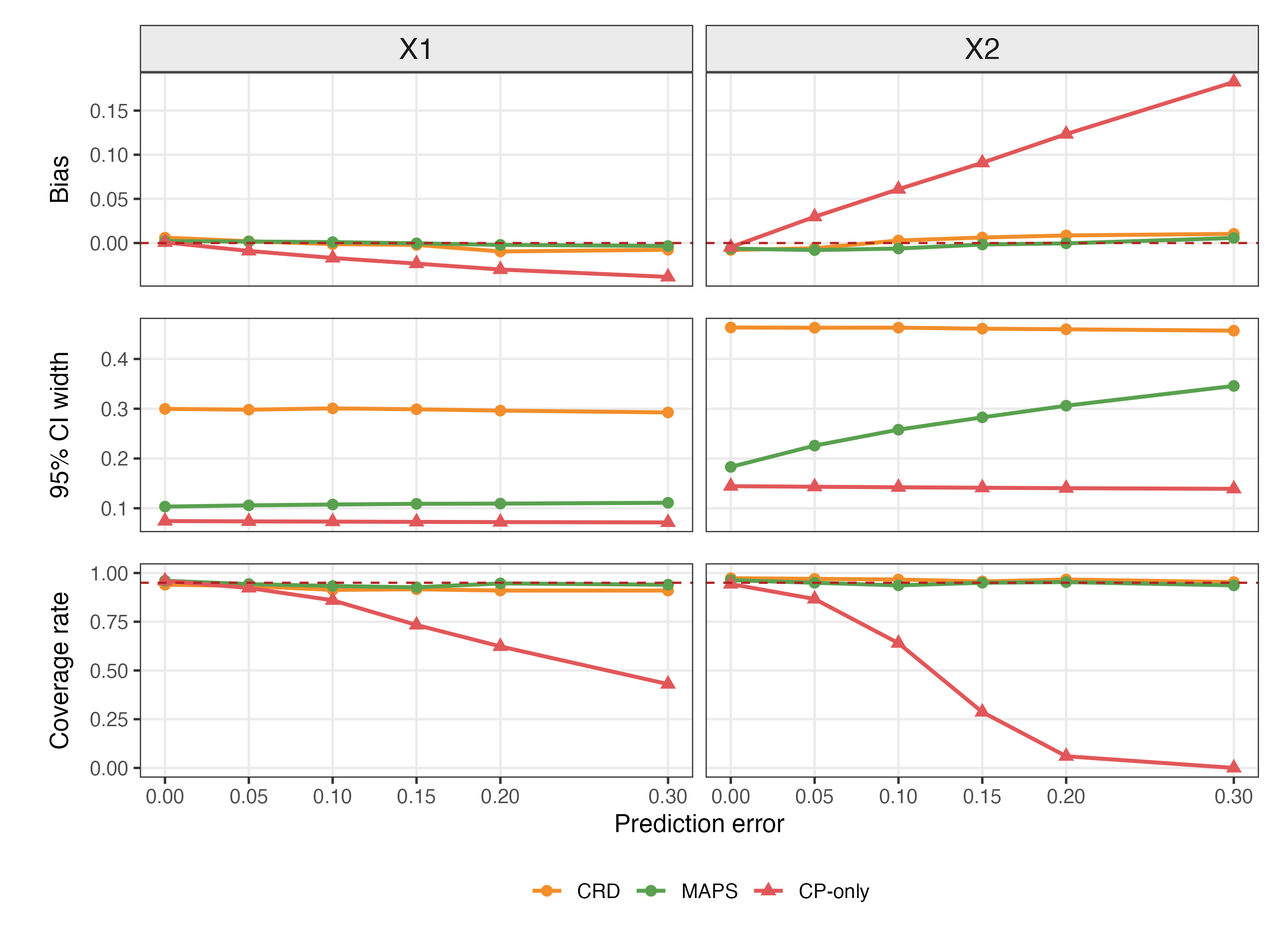}
\caption{Performance across covariate CP prediction-error rates (0\%--30\%), with outcome CP error fixed at $\delta=0$ and $\sigma=0.05$, $N=5000$, 25\% outcome validation, 40\% $x_2$ validation, and 300 replicates per level. Bias and mean CI width are on the log-HR scale. Dashed red lines indicate zero bias and nominal 95\% coverage. CRD, chart-reviewed data analysis.}
\label{fig:sim_4pattern_x2_error}
\end{figure}

\textbf{CP Error in both covariate and outcome.}
Consistent with the outcome-error simulations, MAPS remained approximately unbiased with near-nominal 95\% coverage across all levels of covariate prediction error, while consistently achieving narrower CIs than CRD (Figure~\ref{fig:sim_4pattern_x2_error}). In contrast, CP-only estimates exhibited increasing bias and declining coverage as prediction error increased. Notably, MAPS interval widths for $X_2$ gradually increased as the error rate increased, while remaining substantially narrower than CRD intervals. CP-only estimates showed particularly pronounced bias for $X_2$, reaching approximately 0.18 at the highest error rate, with coverage falling to nearly 0\%. Coverage for $X_1$ also declined to approximately 43\%. These results further demonstrate that MAPS improves precision over CRD and outperforms CP-only estimation.

\section{Real Data Analysis}

To illustrate our MAPS regression method, we analyzed data from a Northwestern Medicine EHR cohort of 1,026 cancer patients who initiated ICI therapy between April 1, 2011, and August 25, 2021. We examined the association between pre-existing AID and the risk of irAEs following ICI treatment. Error-prone clinical proxies for AID history and irAE outcomes were derived from diagnosis codes. Both AID history and irAE outcomes were clinically adjudicated through chart review for all 1,026 patients, providing a unique opportunity to evaluate MAPS under varying proportions of patients with validated labels.

The outcome was time from ICI initiation to the first irAE in a clinical category not documented before treatment. Follow-up ended at the earlier of the last recorded follow-up and 2 years. In addition to pre-existing AID as the primary exposure, the regression model included ICI class, age, and sex to adjust for potential confounding, with PD-1-only therapy as the reference regimen. We used 30-day time intervals and modeled the baseline hazard using a natural cubic spline with three degrees of freedom.

The outcome CP used post-ICI ICD diagnosis codes mapped to irAE categories, excluding categories in the same adjudicated pretreatment history used for the validated outcome. The first qualifying code defined the CP event time; patients without a qualifying code were censored at the end of follow-up. The AID CP was positive if at least one autoimmune ICD-10 diagnosis code was recorded during the 2 years before ICI initiation.

We masked validated irAE outcome and AID history labels at prespecified rates, with validation probabilities dependent on age, sex, and ICI class. For each configuration, we repeated the masking process 100 times and averaged the results across replicates. Both MAPS and CRD used stabilized inverse-probability sampling weights. We estimated log-HRs and their 95\% confidence intervals using MAPS, CRD, and CP-only analyses, with the fully validated analysis serving as the benchmark. Details on masking rates, sampling weights, and implementation are provided in Supplementary Methods, Section~S11.

In the full cohort, the mean age was 66.9 years (SD, 12.4), and 485 patients (47.3\%) were female. Of these patients, 668 (65.1\%) received PD-1-only therapy, 180 (17.5\%) received PD-L1-only therapy, and 178 (17.3\%) received a CTLA-4-containing or combination regimen. Chart review confirmed irAEs in 504 patients (49.1\%) and pre-existing AID in 80 (7.8\%). Additional cohort characteristics are provided in Supplementary Table~S1.

CP-derived labels agreed with chart-reviewed labels in 58.8\% of patients for irAE occurrence and 92.2\% for pre-existing AID. The sensitivity and specificity of CPs were 85.1\% and 33.3\%, respectively, for irAE occurrence, compared with 26.3\% and 97.8\% for AID. These findings indicate substantial misclassification in both CPs, with particularly low specificity for irAEs and low sensitivity for AID. These measures describe binary label classification rather than agreement in event times.

In the analysis using all adjudicated AID and irAE data, AID showed no clear association with irAE risk (HR = 0.93; 95\% CI, 0.67--1.29), whereas the CP-only analysis suggested a significant positive association (HR = 1.53; 95\% CI, 1.12--2.09). Estimates for all other covariates are presented in Supplementary Table~S2.

\begin{figure*}[t]
\centering
\includegraphics[width=0.90\textwidth]{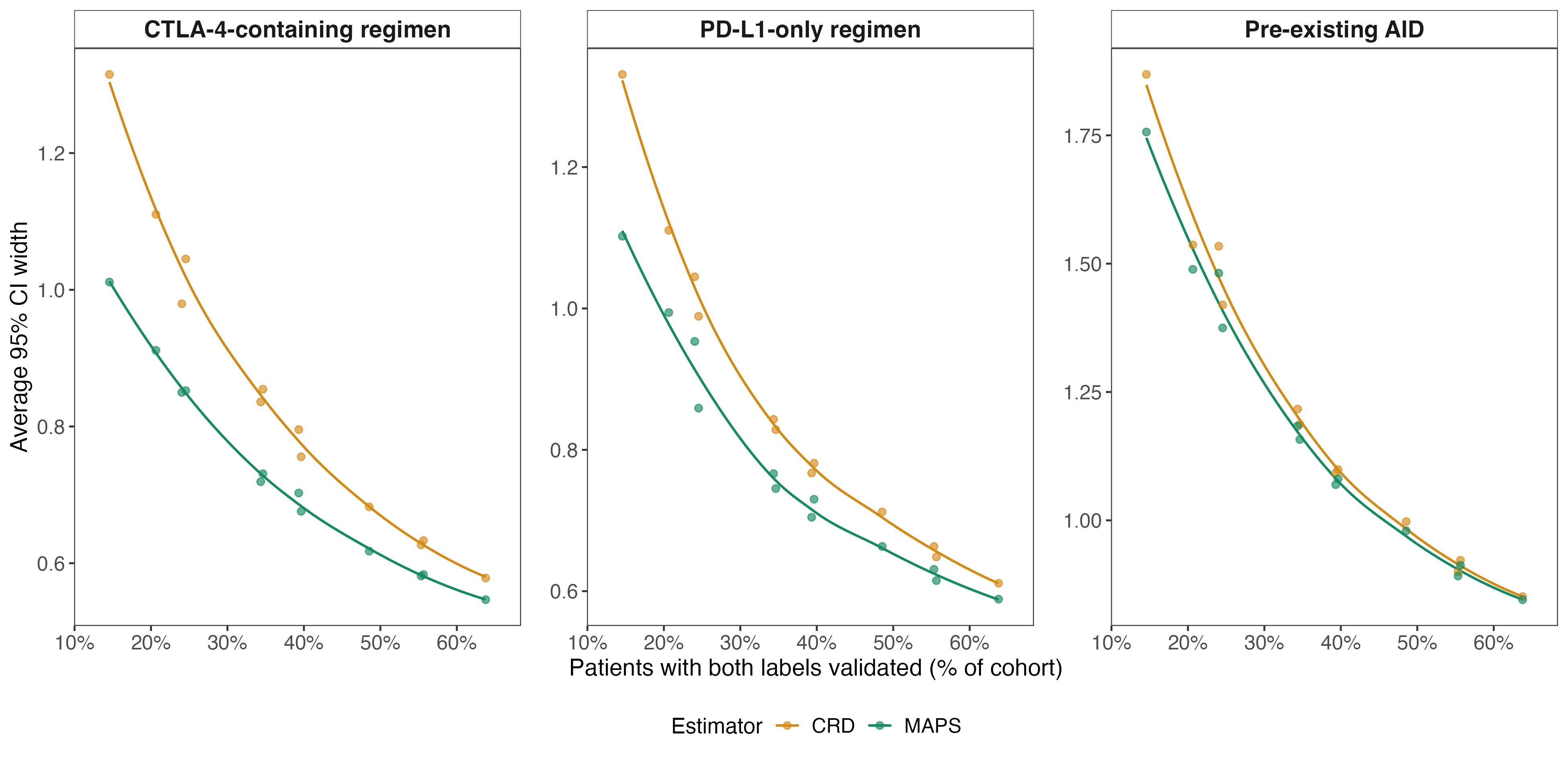}
\caption{Mean 95\% CI widths for MAPS and CRD across validation levels. Points summarize 100 masking replicates per configuration; curves are smoothed guides. The horizontal axis is the mean number of patients with both outcome and AID labels validated divided by the cohort size ($N=1026$), expressed as a percentage (approximately 15\%--64\%). Widths are on the log-HR scale. Both methods used stabilized inverse-probability sampling weights. CRD, chart-reviewed data analysis.}
\label{fig:real_mar_width}
\end{figure*}

As shown in Figure~\ref{fig:real_mar_forest}, MAPS produced mean log-HR estimates closely aligned with those from the fully validated cohort (benchmark) across the three selected covariates and validation proportions. MAPS consistently achieved narrower 95\% CIs than CRD, with CI widths decreasing as the validation proportion increased. The precision gain over CRD was most pronounced when only 15\% of patients were fully validated but diminished as the validation proportion increased to 56\% (Figure~\ref{fig:real_mar_width}). Substantial errors in CP-derived labels may have limited the additional information gained from incorporating them. At approximately 15\% validation, mean CI widths were 23.1\% lower for CTLA-4-containing regimens and 6.0\% lower for pre-existing AID than with CRD. In contrast, CP-only estimates showed substantial deviations from the benchmark for CTLA-4-containing regimens and pre-existing AID. Overall, MAPS improved precision over CRD while closely reproducing the fully validated estimates, particularly when validation data were limited. Results for all model covariates are provided in Supplementary Table~S3.

\begin{figure*}[t]
\centering
\includegraphics[width=0.95\textwidth]{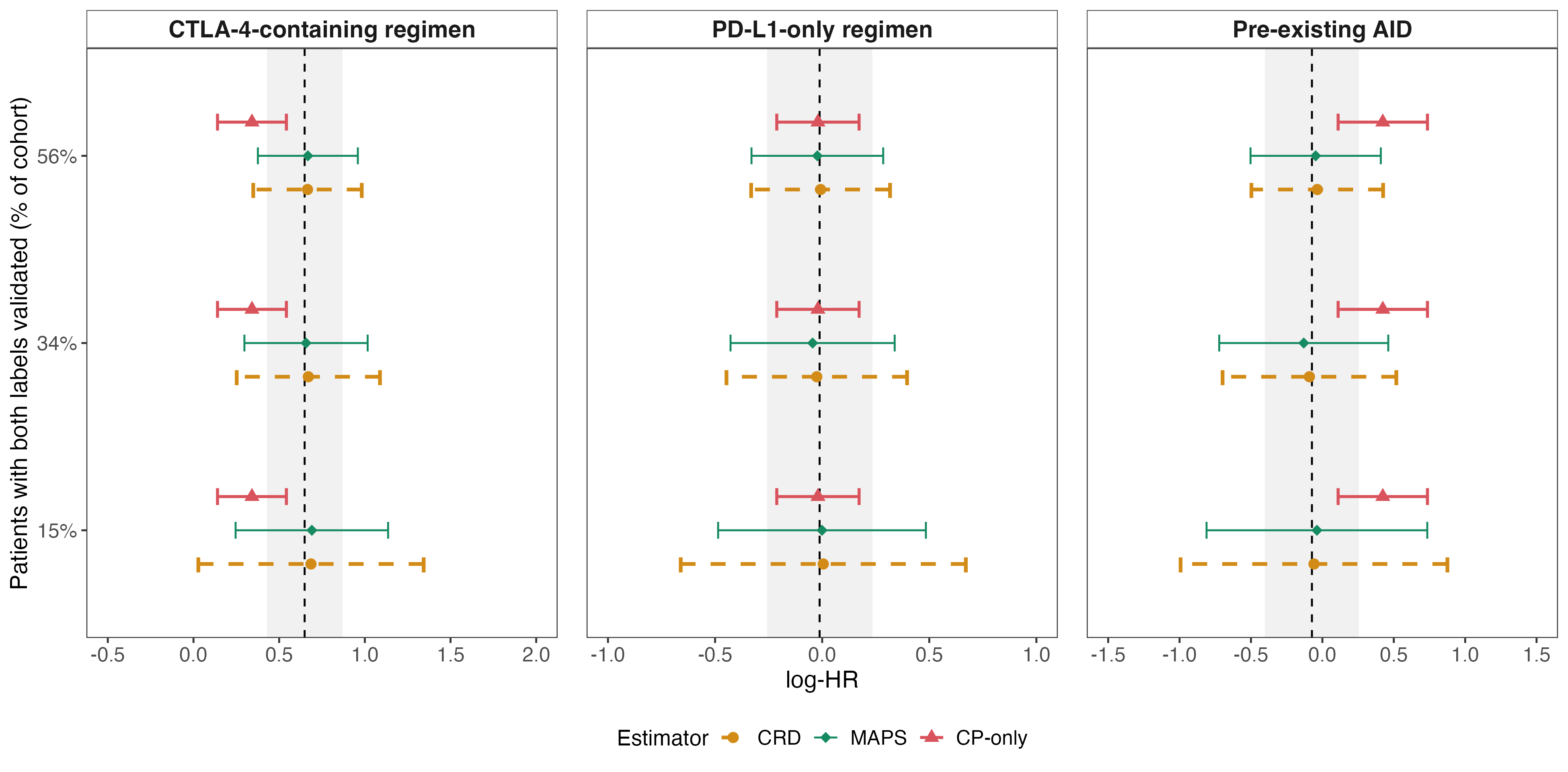}
\caption{Masked-validation estimates for time to an irAE. Mean log-HRs and mean lower and upper 95\% CI limits are shown across 100 masking replicates per setting. Percentages indicate the proportion of the cohort with both validated survival outcomes and AID history available after masking, corresponding to $\mathcal D\equiv\mathcal M_0$. Reference groups are PD-1-only therapy and no pre-existing AID. The vertical dashed black line and gray band show the estimate and 95\% CI from the fully validated cohort, respectively. CRD, chart-reviewed data analysis.}
\label{fig:real_mar_forest}
\end{figure*}

\section{Discussion}

We developed MAPS regression to address phenotyping errors in EHR-based survival association studies when both outcomes and covariates are error-prone and chart-reviewed labels are available for only a subset of patients. MAPS extends prediction-powered inference to censored time-to-event outcomes using discrete-time survival regression with a complementary log-log link. Through a multivariate augmentation framework, MAPS combines validated and CP-derived labels across different validation patterns, using data-adaptive weights to improve statistical efficiency over CRD analysis, which relies solely on fully validated records, while preserving consistency under appropriate assumptions. Robust sandwich variance estimation accounts for within-patient dependence across follow-up intervals, while inverse-probability sampling weights accommodate covariate-dependent chart-review selection.

Our main simulation studies demonstrated that MAPS substantially improved precision over CRD while maintaining negligible bias and generally near-nominal 95\% confidence interval coverage. These findings held when CP errors affected the survival outcome alone and when both the outcome and a covariate were error-prone. In contrast, CP-only analyses exhibited increasing bias and substantial losses in coverage as CP errors increased, highlighting the importance of using validated labels for bias correction. The precision gains from MAPS depended on CP quality. As CP errors increased, MAPS confidence intervals widened but remained narrower than those from CRD. These findings demonstrate that MAPS can improve efficiency even with imperfect CPs, although the magnitude of improvement depends on how strongly CP-derived labels are associated with the true phenotypes.  

The real-data analysis showed that substantial errors in the AID and irAE CPs may have limited the precision gains from MAPS. These gains further diminished as validation rates increased, because more information was already available from validated records, reducing the additional benefit of CP-derived labels. These findings highlight the importance of both CP quality and validation rates in determining the efficiency gains from MAPS.

When chart review is conducted for two variables, four distinct validation patterns are possible, as illustrated in Figure~\ref{fig:debias}. However, in practice, only a subset of these patterns may be observed. For example, when both variables are validated for the same patients, only fully validated ($\mathcal M_0$) and fully unvalidated ($\mathcal M_3$) records are available. We therefore conducted additional simulation studies to evaluate the performance of MAPS under different validation-pattern configurations. When one variable was validated for half of the patients outside the fully validated subset, incorporating these labels reduced CI widths by 8.3\% and 20.4\% for the two regression coefficients compared with an augmentation approach that did not use this additional validation information (Supplementary Figure~S12). The gains were not uniform across coefficients and validation allocations.

Several limitations warrant consideration. When chart-review selection depends on patient characteristics, valid inference requires positive validation-pattern probabilities, a correctly specified selection model, and corresponding inverse-probability sampling weights. In addition, small fully validated samples may lead to unstable covariance estimates and data-adaptive augmentation weights, potentially affecting confidence interval coverage. For example, in simulations with only 10\% outcome validation, MAPS empirical coverage ranged from 91\% to 95\%, slightly below the nominal level in some settings (Supplementary Table~S6). Future work will evaluate MAPS across multiple healthcare systems to assess its generalizability. Integrating MAPS with targeted chart-review sampling strategies, which use CP-derived labels to preferentially select informative patients for validation, such as patients whose CPs indicate the event of interest \citep{yin2022cost}, may further improve statistical efficiency while reducing chart-review costs.

\section{Conclusion}
MAPS addresses phenotyping errors in EHR-based survival studies by combining limited chart-reviewed data with error-prone CPs across different validation patterns. Simulations and an analysis of real EHR data with masked validation labels demonstrated improved precision over CRD analysis, with approximately unbiased estimates and generally near-nominal confidence interval coverage in the main simulations. Although efficiency gains depend on CP quality and validation rates, MAPS offers a practical approach to improving EHR-based survival inference with limited chart-review resources.

\bibliographystyle{oup-abbrvnat}
\bibliography{psppi_survival_jamia_refs}

\end{document}


\begin{center}
{\large\bfseries Multivariate Augmented Prediction-Powered Survival Inference for EHR Association Studies With Limited Chart Review\par}
\medskip
{\small Xicheng Xie, Katherine Fiona Simpson, Hyojung Jang, Theresa L. Walunas, Lili Zhao\par}
\end{center}

\section*{Supplementary Methods}

\subsection{Grouped-Time Cloglog Model From Continuous-Time Proportional Hazards}
\label{supp:s1}
The main paper uses a grouped-time complementary log-log model because it fits naturally into a person-period representation while retaining the proportional hazards interpretation under the grouped-time model.\citep{prentice1978grouped,guo1994multivariate,carlin1999discrete} The derivation is included here to keep the main Methods section focused on the estimator.

Assume a continuous-time proportional hazards model
\[
h(t\mid X)=h_0(t)\exp(\beta^\top X),
\]
with baseline hazard $h_0(t)$ and cumulative baseline hazard $H_0(t)=\int_0^t h_0(s)\,ds$. The corresponding survival function is
\[
S(t\mid X)=\exp\{-\exp(\beta^\top X)\,H_0(t)\}.
\]
Let $0=t_0<t_1<\cdots<t_J$ be interval boundaries and define the grouped-time hazard
\[
p_{ij}
=
\Pr\{T_i\in(t_{j-1},t_j]\mid T_i>t_{j-1},X_i\}.
\]
Then
\[
p_{ij}
=
1-\frac{S(t_j\mid X_i)}{S(t_{j-1}\mid X_i)}
=
1-\exp\!\left[-\exp(\beta^\top X_i)\{H_0(t_j)-H_0(t_{j-1})\}\right].
\]
If $H_{0j}=H_0(t_j)-H_0(t_{j-1})$ denotes the baseline cumulative hazard increment on interval $j$, then
\[
\log\{-\log(1-p_{ij})\}
=
\log(H_{0j})+\beta^\top X_i.
\]
Thus the grouped-time model
\[
\log\{-\log(1-p_{ij})\}
=
\alpha_j+\beta^\top X_i,
\qquad
\alpha_j=\log(H_{0j}),
\]
retains the proportional hazards interpretation for the covariate coefficients. In the manuscript, the baseline term is represented by a flexible basis in time rather than a separate parameter for each interval. Changes in interval width change the baseline increments represented by this time function, while the covariate coefficients retain their log hazard ratio (log-HR) interpretation under the model.

We also compared fitted cloglog and Cox coefficients in 200 simulated cohorts of $N=5000$ subjects per setting. Event times followed a Weibull proportional hazards model with shape 1.5, baseline scales of 12, 24, or 48, and covariate hazard ratios of 1.4 and 0.75. The two standard-normal covariates had correlation 0.2. An exponential censoring rate was calibrated to 35\% censoring at scale 24 and held fixed across settings. Cloglog models used interval widths of 1, 3, or 6 and a 3-degree-of-freedom natural-spline baseline time effect.

\begin{figure}[!htbp]
\centering
\includegraphics[width=\linewidth]{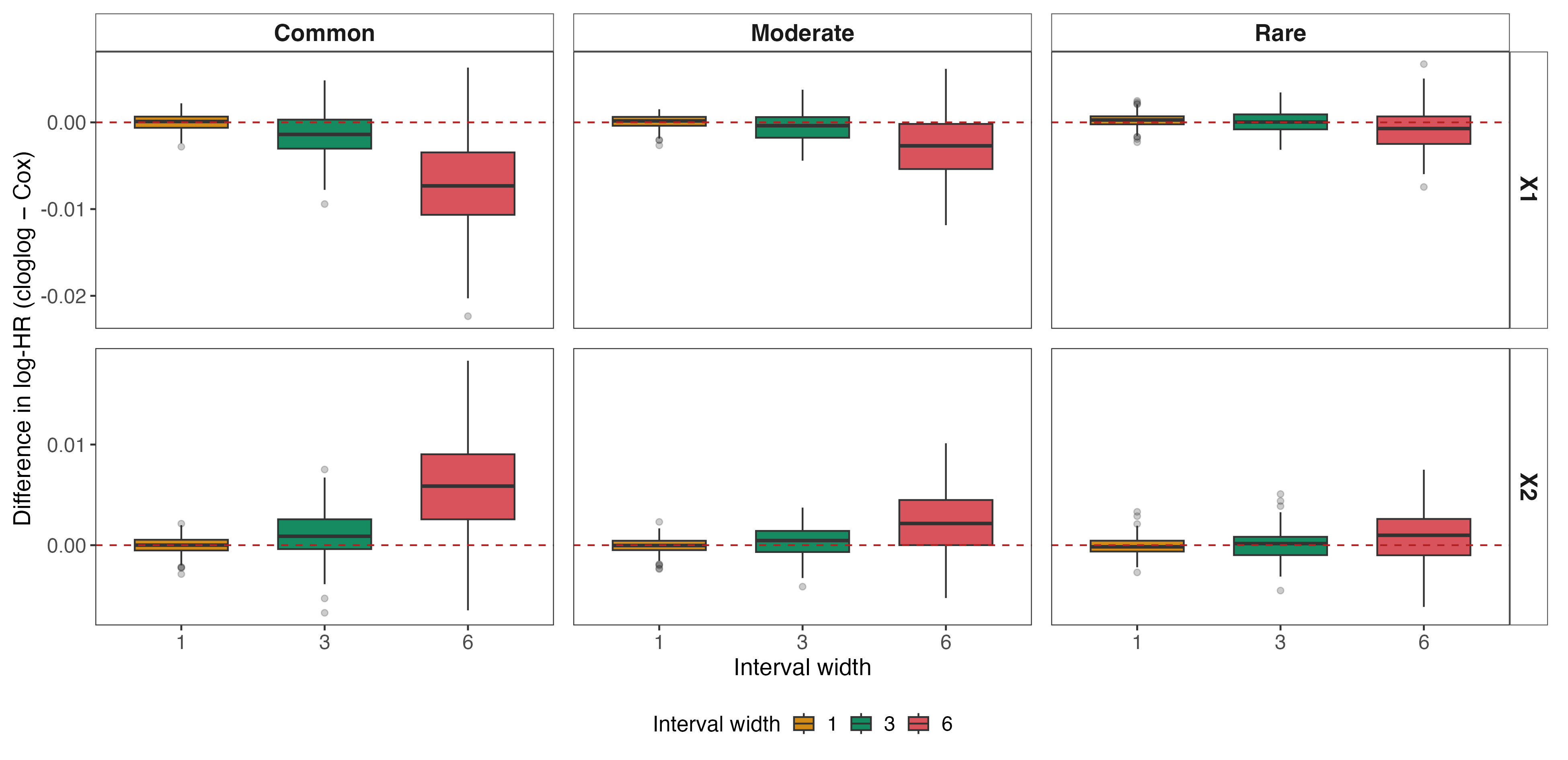}
\caption{Differences between cloglog and Cox log-HR estimates across interval widths. Each setting used 200 simulated cohorts of 5000 subjects. Columns correspond to Weibull baseline scales of 12, 24, and 48, with mean observed event proportions of 79.5\%, 64.9\%, and 45.9\%, respectively; rows correspond to the two covariates. The cloglog model used a 3-degree-of-freedom natural-spline baseline time effect. The dashed line indicates equal estimates.}
\label{fig:s-cloglog-cox}
\end{figure}

The estimates were closely aligned, with smaller differences at narrower interval widths (Figure~\ref{fig:s-cloglog-cox}). The grouped-time identity above is exact for interval-specific baseline increments; the fitted spline baseline and terminal censoring-interval convention can produce small differences between the estimates.

\subsection{General $K$-Pattern Notation and Masking Formalism}
\label{supp:s2}
For subject $i=1,\ldots,N$, define the full subject-level survival object as
\[
\mathcal S_i^{\mathrm{full}}=(X_i,T_i,C_i,\Delta_i),
\qquad
U_i=\min(T_i,C_i),
\qquad
\Delta_i=\ind\{T_i\le C_i\}.
\]
After grouped-time expansion, the corresponding long-format survival object is
\[
\mathcal D_i^{\mathrm{full}}=(X_i,d_i,J_i),
\qquad
d_i=(d_{i1},\ldots,d_{iJ_i}),
\]
where $d_i$ is the subject-level event path in person-period form and $J_i$ is the number of rows implied by follow-up. Let
\[
M_i\in\{\infty,1,\ldots,K\}
\]
denote the realized missingness pattern, with $M_i=\infty$ for fully observed subjects and $k=1,\ldots,K$ for incomplete patterns. Each pattern may involve missing event-time outcomes, missing baseline covariates, or both.

For each incomplete pattern $k$, let $\mathcal R_k$ denote the coarsening operator that removes exactly the components not observed under pattern $k$. The observed dataset is
\[
\mathcal D_{\mathrm{obs}}
=
\{(\mathcal R_{M_i}(\mathcal D_i^{\mathrm{full}}),M_i):i=1,\ldots,N\}.
\]
This formulation accommodates arbitrary non-monotone missingness patterns, provided pattern membership is defined at the subject level. A four-pattern survival implementation is a special case of this notation.

\subsection{Subject-Level Completion Operators and Pseudo-Trajectories}
\label{supp:s3}
Let $\mathcal A_k$ denote the pattern-specific learner-generated completion rule. Applying $\mathcal A_k$ after masking produces the completed subject-level object
\[
\widetilde{\mathcal D}_i^{(k)}
=
\mathcal A_k\{\mathcal R_k(\mathcal D_i^{\mathrm{full}})\}
=
(\widetilde X_i^{(k)},\widetilde d_i^{(k)},J_i^{\ast(k)}).
\]
The same rule $\mathcal A_k$ is used in two places: complete cases are artificially masked to resemble pattern $k$ and then completed, while subjects actually observed under pattern $k$ are completed with the same rule. The completed values are used to form estimating equations and rectifiers; they are not treated as ground truth.

Completion is performed at the subject level and the result is then expanded into person-period form. This order avoids impossible row-level trajectories, such as more than one event row for a single subject. A completed trajectory may have a different number of intervals from the observed trajectory, so $J_i^{\ast(k)}$ need not equal $J_i$. With the same time basis $b(t)$ used in the main text, the completed row design is
\[
\widetilde z_{ij}^{(k)}
=
\begin{bmatrix}
b((j-0.5)dt)\\
\widetilde X_i^{(k)}
\end{bmatrix},
\qquad j=1,\ldots,J_i^{\ast(k)}.
\]

\subsection{Row-Level Score and Subject-Level Aggregation}
\label{supp:s4}
Let $b(t)$ be the common time basis and define
\[
z_{ij}=
\begin{bmatrix}
b(t_{ij}^{\mathrm{mid}})\\
X_i
\end{bmatrix},
\qquad
\vartheta=
\begin{bmatrix}
\alpha\\
\beta
\end{bmatrix},
\qquad
p(\vartheta;z)=1-\exp\{-\exp(z^\top\vartheta)\}.
\]
Here $\vartheta$ denotes a generic parameter vector with the same structure as $\theta$, $\gamma_{1,k}$, or $\gamma_{2,k}$. For a Bernoulli outcome $y$ and design vector $z$, the complementary log-log row-level score is
\[
s(\vartheta;y,z)
=
z\,\exp(z^\top\vartheta)\,
\frac{y-p(\vartheta;z)}{p(\vartheta;z)}.
\]
For a subject with row count $J_i^\circ$, outcome path $y_i^\circ=(y_{i1}^\circ,\ldots,y_{iJ_i^\circ}^\circ)$, and design rows $z_i^\circ=(z_{i1}^\circ,\ldots,z_{iJ_i^\circ}^\circ)$, the subject-level cluster score is
\[
S_i(\vartheta;y_i^\circ,z_i^\circ,J_i^\circ)
=
\sum_{j=1}^{J_i^\circ} s(\vartheta;y_{ij}^\circ,z_{ij}^\circ).
\]
The observed-data anchor score is
\[
S_i^\theta(\theta)=S_i(\theta;d_i,z_i,J_i).
\]
For pattern $k$, the completed-data cluster score is
\[
S_i^{(k)}(\gamma)
=
S_i\!\left(\gamma;\widetilde d_i^{(k)},\widetilde z_i^{(k)},J_i^{\ast(k)}\right).
\]

\subsection{Bread and Meat Contributions for the Cloglog Estimating Equations}
\label{supp:s5}
The implementation uses the expected information for the bread and patient-level estimating functions for the meat. For the cloglog model, the expected row-level information is defined as follows. For a row with design vector $z$ and parameter $\vartheta$, define
\[
\mathcal I_\eta(\vartheta;z)
=
\frac{\exp\{2z^\top\vartheta\}\,[1-p(\vartheta;z)]}{p(\vartheta;z)}.
\]
The row-level bread contribution is
\[
B(\vartheta;z)
=
z z^\top\,\mathcal I_\eta(\vartheta;z).
\]
For the anchor, the subject-level bread contribution is
\[
B_{\theta,i}(\theta)
=
w_i^{\mathrm{cc}}\sum_{j=1}^{J_i} B(\theta;z_{ij}),
\]
with analogous expressions for each pattern-$k$ estimator using the completed rows $\widetilde z_{ij}^{(k)}$ and the corresponding weights.

For a parameter block $a$, let $U_{a,i}$ denote the patient-level weighted estimating function and $B_{a,i}$ the corresponding information contribution. Define
\[
B_a=\frac{1}{N}\sum_{i=1}^N B_{a,i}.
\]
For parameter blocks $a$ and $b$, the cluster meat is
\[
M_{a,b}=\frac{1}{N}\sum_{i=1}^N U_{a,i}U_{b,i}^\top.
\]
The corresponding subject-level sandwich block is
\[
\widehat\Sigma_{a,b}
=
\frac{1}{N}B_a^{-1}M_{a,b}B_b^{-\top}.
\]

\subsection{Detailed MAPS Estimating Equations With Propensity Weights}
\label{supp:s6}
Under MAR, let
\[
\pi_r(O_i)=\Pr(M_i=r\mid O_i),
\qquad r\in\{\infty,1,\ldots,K\},
\]
where $O_i$ is a common observed-data vector available for every subject.\citep{sun2018inverse} The variables in $O_i$ may overlap with $X_i$, but any variable in $O_i$ must be observed for all subjects. These probabilities may be estimated via multinomial logistic regression or another prespecified missingness model and yield the inverse-probability weights
\[
w_i^{\mathrm{cc}}=\frac{\ind\{M_i=\infty\}}{\hat\pi_\infty(O_i)},
\qquad
w_i^{(k)}=\frac{\ind\{M_i=k\}}{\hat\pi_k(O_i)},
\qquad k=1,\ldots,K.
\]
Under MCAR these reduce to corresponding pattern-frequency constants. Stabilized or truncated versions of these weights can be used in sensitivity analyses when estimated pattern probabilities are close to zero.

The weighted estimating equations are
\[
U_\theta(\theta)
=
\frac{1}{N}\sum_{i=1}^N
w_i^{\mathrm{cc}}\,S_i^\theta(\theta)=0,
\]
\[
U_{\gamma_{1,k}}(\gamma_{1,k})
=
\frac{1}{N}\sum_{i=1}^N
w_i^{\mathrm{cc}}\,S_i^{(k)}(\gamma_{1,k})=0,
\]
\[
U_{\gamma_{2,k}}(\gamma_{2,k})
=
\frac{1}{N}\sum_{i=1}^N
w_i^{(k)}\,S_i^{(k)}(\gamma_{2,k})=0.
\]
The pattern-specific rectifier is
\[
\Delta_k=\gamma_{1,k}-\gamma_{2,k},
\qquad
\hat\Delta_k=\hat\gamma_{1,k}-\hat\gamma_{2,k},
\]
and the summed rectifier is
\[
\Delta_{\mathrm{sum}}=\sum_{k=1}^K \Delta_k,
\qquad
\hat\Delta_{\mathrm{sum}}=\sum_{k=1}^K \hat\Delta_k.
\]
With a common empirical weight matrix, the MAPS estimator is
\[
\hat\theta_{\mathrm{MAPS}}
=
\hat\theta-\hat W\hat\Delta_{\mathrm{sum}}.
\]
The broader asymptotic theory follows the general PS-PPI development for Z-estimation problems and applies here after replacing the generic score with the subject-level cloglog cluster score.\citep{chen2025generalmissing}

\subsection{Subject-Level Sandwich Covariance and Expanded Cross-Block Formulas}
\label{supp:s7}
All covariance calculations use patients as independent clusters. Interval contributions are summed within each patient before forming the outer products, retaining dependence across follow-up intervals \citep{liang1986longitudinal}. The relevant patient-level estimating functions are
\[
U_{\theta,i}(\theta)=w_i^{\mathrm{cc}}S_i^\theta(\theta),
\qquad
U_{\gamma_{1,k},i}(\gamma_{1,k})=w_i^{\mathrm{cc}}S_i^{(k)}(\gamma_{1,k}),
\]
\[
U_{\gamma_{2,k},i}(\gamma_{2,k})=w_i^{(k)}S_i^{(k)}(\gamma_{2,k}).
\]
Using the row-level bread formula from Supplementary Section~\ref{supp:s5}, the patient-level contribution to the information sum defining $B_\theta$ is $w_i^{\mathrm{cc}}\sum_j B(\theta; z_{ij})$, with analogous expressions for $B_{\gamma_{1,k}}$ and $B_{\gamma_{2,k}}$. The cross-meat $\frac{1}{N}\sum_i U_{a,i}U_{b,i}^\top$ is nonzero when the two estimating functions share nonzero subject contributions. For example, $U_{\theta,i}$ and $U_{\gamma_{1,k},i}$ are correlated for complete-case subjects because both depend on $w_i^{\mathrm{cc}}$.

The cross-blocks required for the weight matrix and MAPS covariance are
\[
\widehat\Sigma_{\theta,\Delta}
=
\sum_{k=1}^K
\left(
\widehat\Sigma_{\theta,\gamma_{1,k}}
-\widehat\Sigma_{\theta,\gamma_{2,k}}
\right),
\]
and
\[
\widehat\Sigma_\Delta
=
\sum_{k=1}^K\sum_{k'=1}^K
\widehat{\operatorname{Cov}}(\hat\Delta_k,\hat\Delta_{k'}),
\]
where
\[
\widehat{\operatorname{Cov}}(\hat\Delta_k,\hat\Delta_{k'})
=
\widehat\Sigma_{\gamma_{1,k},\gamma_{1,k'}}
-\widehat\Sigma_{\gamma_{1,k},\gamma_{2,k'}}
-\widehat\Sigma_{\gamma_{2,k},\gamma_{1,k'}}
+\widehat\Sigma_{\gamma_{2,k},\gamma_{2,k'}}.
\]

\subsection{Optimal or Empirical Weight Matrix}
\label{supp:s8}
At the population level, define
\[
\Sigma_{\theta,\Delta}
=
\operatorname{Cov}(\hat\theta,\hat\Delta_{\mathrm{sum}}),
\qquad
\Sigma_\Delta
=
\operatorname{Var}(\hat\Delta_{\mathrm{sum}}),
\qquad
\Sigma_{\Delta,\theta}=\Sigma_{\theta,\Delta}^\top.
\]
For a generic common weight matrix $W$, the covariance of $\hat\theta-W\hat\Delta_{\mathrm{sum}}$ is
\[
\Sigma_{\mathrm{MAPS}}(W)
=
\Sigma_\theta
-W\Sigma_{\Delta,\theta}
-\Sigma_{\theta,\Delta}W^\top
+W\Sigma_\Delta W^\top.
\]
Minimizing this quadratic form gives the common population weight
\[
W=\Sigma_{\theta,\Delta}\Sigma_\Delta^{-1}.
\]
The empirical version used in the estimator is
\[
\hat W
=
\widehat\Sigma_{\theta,\Delta}\widehat\Sigma_\Delta^{-1},
\]
with covariance estimate
\[
\widehat\Sigma_{\mathrm{MAPS}}
=
\widehat\Sigma_\theta
-\widehat\Sigma_{\theta,\Delta}
\widehat\Sigma_\Delta^{-1}
\widehat\Sigma_{\theta,\Delta}^\top.
\]
The corresponding population form is
\[
\Sigma_{\mathrm{MAPS}}
=
\Sigma_\theta
-\Sigma_{\theta,\Delta}\Sigma_\Delta^{-1}\Sigma_{\Delta,\theta}
\;\preceq\;
\Sigma_\theta
\]
when the population weight is used. Equality occurs when the summed rectifier carries no covariance information about $\hat\theta$. These covariance estimates use the information bread defined in Section~\ref{supp:s5} and treat fitted CP and validation models as fixed. They do not propagate uncertainty from fitting those models. Confidence-interval calibration is therefore assessed empirically in the reported simulations.

\subsection{Simulation Settings and Evaluation}
\label{supp:s9}
All simulations reported in the main text used $N=5000$, 300 replicates per prediction-noise or covariate-error level, Weibull shape $1.5$, baseline scale $3$, and covariate coefficients $\beta_1=\log(1.4)$ and $\beta_2=\log(0.75)$. Exponential censoring was calibrated to approximately 35\%. Analyses used intervals of width $dt=0.5$ and a common 3-degree-of-freedom natural-spline basis for baseline time. The covariate distributions and validation-selection models are specified in the main Methods. Stabilized inverse-probability weights used a lower pattern-probability floor of $0.01$.

\paragraph{Outcome prediction-noise analysis.}
Both covariates were observed without error, and outcome validation targeted 10\% of subjects. CP-derived event times used the linear predictor
\[
\widetilde\ell_i=X_i^\top\beta+\varepsilon_i,
\qquad\varepsilon_i\overset{\mathrm{ind}}{\sim}N(\delta,\sigma^2),
\]
with $\delta=0.30$ and $\sigma\in\{0,0.10,0.20,0.30,0.50,0.75,1.00\}$. Errors were independent of the covariates. For a shared latent draw $E_i\sim\operatorname{Exp}(1)$, true and CP-derived event times were
\[
T_i=3\{E_i\exp(-X_i^\top\beta)\}^{1/1.5},
\qquad
\widetilde T_i=3\{E_i\exp(-\widetilde\ell_i)\}^{1/1.5}.
\]
They also shared the same censoring time. With $\sigma=0$, the fixed shift rescales the baseline hazard and does not change the conditional covariate hazard ratios. This cell therefore has a mean prediction shift even though it has no random prediction noise. Random seeds were paired across noise levels, preserving the true data and validation indicators. Each of the seven cells contained 300 successful replicates, with no failed attempts.

Table~\ref{tab:s6-simulation-calibration} compares empirical standard deviations with mean estimated standard errors for MAPS and chart-reviewed data (CRD) analysis. Mean MAPS standard errors were smaller than empirical standard deviations at some noise levels, consistent with coverage below 95\% in those settings. Monte Carlo standard errors are reported to show the uncertainty in the empirical coverage estimates.

\begin{figure}[!htbp]
\centering
\includegraphics[width=\linewidth]{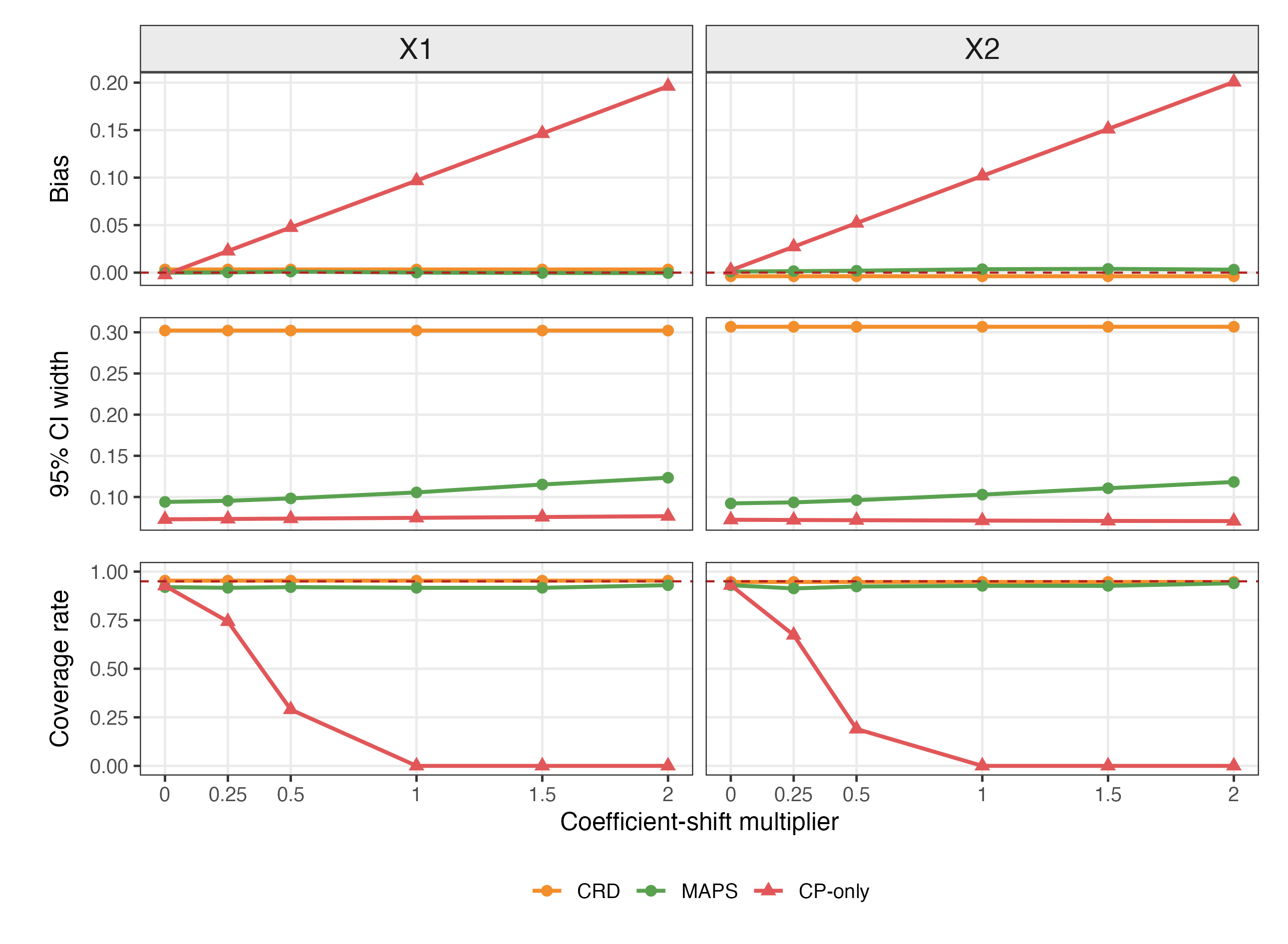}
\caption{Performance with systematic coefficient shifts in the same direction in the outcome CP. The CP linear predictor was $X_i^\top\{\beta+m(0.10,0.10)^\top\}+\varepsilon_i$, with $\varepsilon_i\sim N(0,0.10^2)$ and $m=0,0.25,0.50,1,1.50,2$. Each setting used 300 replicates with $N=5000$ and 10\% EHR-guided outcome validation. Bias and mean 95\% confidence interval width are on the log-HR scale. Dashed red lines indicate zero bias and nominal 95\% coverage.}
\label{fig:s-coefficient-shift-same}
\end{figure}

\begin{figure}[!htbp]
\centering
\includegraphics[width=\linewidth]{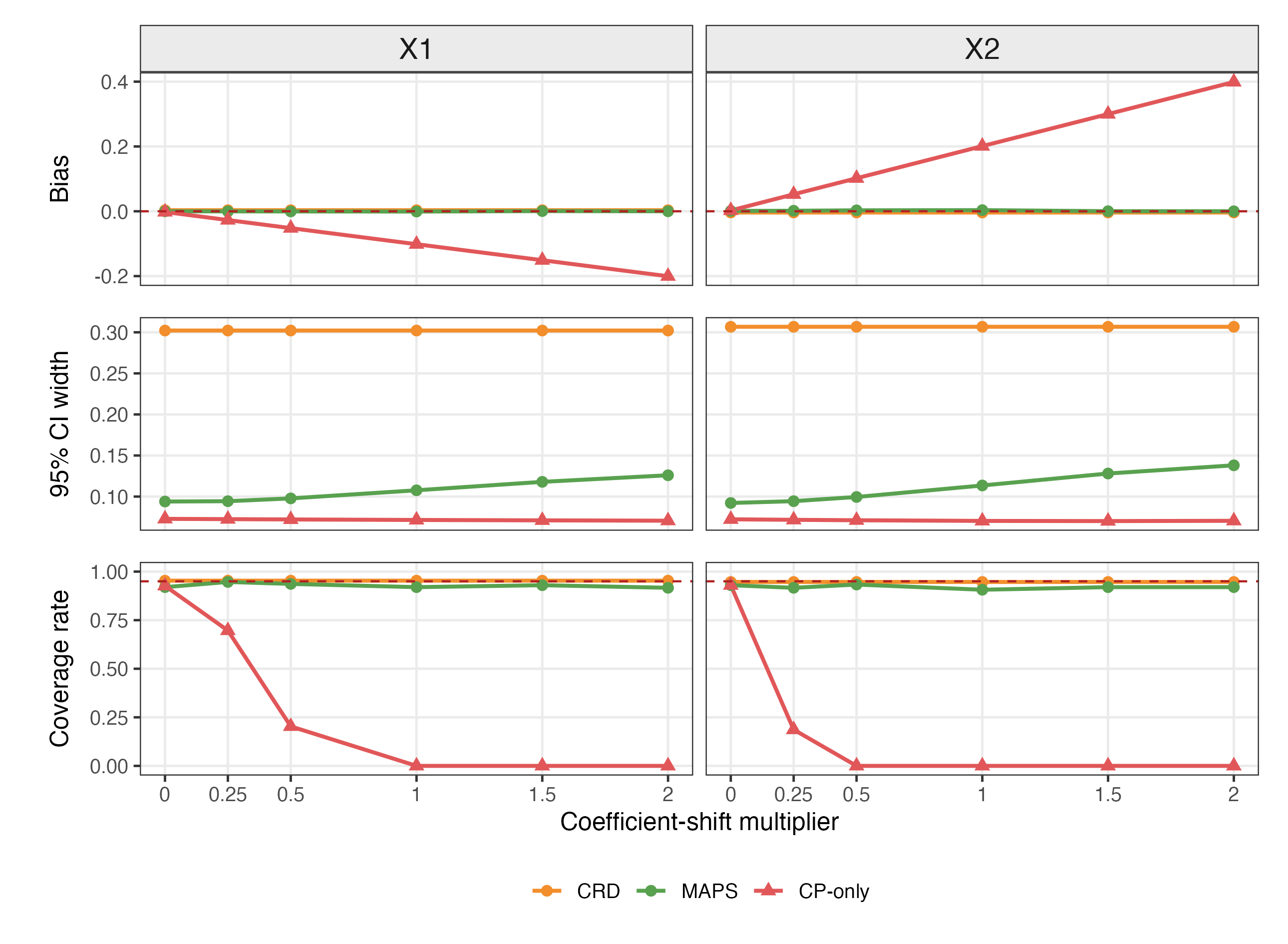}
\caption{Performance with systematic coefficient shifts in opposite directions in the outcome CP. The CP linear predictor was $X_i^\top\{\beta+m(-0.10,0.20)^\top\}+\varepsilon_i$, with $\varepsilon_i\sim N(0,0.10^2)$ and $m=0,0.25,0.50,1,1.50,2$. Each setting used 300 replicates with $N=5000$ and 10\% EHR-guided outcome validation. Bias and mean 95\% confidence interval width are on the log-HR scale. Dashed red lines indicate zero bias and nominal 95\% coverage.}
\label{fig:s-coefficient-shift-opposite}
\end{figure}

\paragraph{Systematic coefficient error in the outcome CP.}
To examine systematic error that changes covariate effects rather than only the baseline hazard, we also generated CP-derived event times using
\[
\widetilde\ell_i=X_i^\top(\beta+md)+\varepsilon_i,
\qquad\varepsilon_i\overset{\mathrm{ind}}{\sim}N(0,0.10^2).
\]
We used $d=(0.10,0.10)^\top$ or $d=(-0.10,0.20)^\top$ and varied $m$ over $\{0,0.25,0.50,1,1.50,2\}$. These shifts were fixed within each setting. The cohort size, true event-time model, shared latent draw and censoring time, 10\% EHR-guided outcome validation, and fitting procedures were unchanged. Each setting had 300 successful replicates. MAPS retained small average bias and narrower intervals than CRD as coefficient shifts increased (Figures~\ref{fig:s-coefficient-shift-same} and~\ref{fig:s-coefficient-shift-opposite}). MAPS coverage ranged from 91\% to 94\% for shifts in the same direction and from 91\% to 95\% for shifts in opposite directions, while CRD coverage remained approximately 95\%. Thus, small average bias did not ensure nominal coverage in these settings.

\paragraph{Covariate prediction-error analysis.}
The probability of an incorrect $x_2$ label varied over $\{0,0.05,0.10,0.15,0.20,0.30\}$. Outcome and $x_2$ validation targeted 25\% and 40\%, respectively. Outcome CPs used $\widetilde\ell_i=X_i^\top\beta+\varepsilon_i$ with $\delta=0$ and $\sigma=0.05$, leaving the covariate coefficients unchanged. Each reported method had 300 evaluable replicates per error level.

\paragraph{Contribution of partial validation and fully validated sample size.}
We conducted two additional experiments using the joint-error setting with $N=5000$, outcome CP error $\varepsilon_i\sim N(0,0.05^2)$, and a 10\% probability of an incorrect $x_2$ label. We selected fully validated patients with probability $p_D(x_1)=\operatorname{expit}(c_0+0.6x_1)$, calibrating $c_0$ to the target subset size. Among the remaining patients, a proportion $q$ had one variable validated, split equally between covariate-only and outcome-only validation; the rest had neither variable validated. This controlled MAR allocation differs from the sequential validation scheme in the main simulations.

To assess the contribution of partial labels, we targeted 500 fully validated patients and varied $q$ over $\{0,0.25,0.50,0.75\}$. We compared MAPS with CP augmentation that ignored partially validated labels, labeled MAPS (pooled CPs). The comparator used the same CRD estimate and formed a single rectifier from fits using both CP-derived variables in $\mathcal D$ and its complement $\mathcal A$. We estimated $p_D$ by logistic regression on $x_1$ and the conditional pattern probabilities from the allocation proportions in $\mathcal A$. Stabilized weights used a probability floor of 0.01; pooled-group weights used $1-\widehat p_D$ before flooring.

To assess validation-sample size, we fixed $q=0.50$ and targeted 100, 250, 500, or 1000 fully validated patients. The simulated data, CPs, and validation uniforms were paired across settings. For each replicate, time-basis knots and boundaries were selected once from the full-cohort true and CP follow-up times and held fixed across settings and estimators. All models used $dt=0.5$, a 3-degree-of-freedom natural-spline baseline time effect, and patient-level sandwich covariance estimation conditional on the fitted validation weights. Each setting used 300 attempted replicates, all of which yielded estimates for every method; failed fits were not replaced. The 500-patient, $q=0.50$ setting was shared between the two experiments.

\paragraph{Comparators and performance measures.}
CRD analysis used records with all required validated labels, whereas CP-only analysis substituted CP-derived labels throughout the cohort. MAPS and CRD used the validation weights; CP-only analysis used unit weights. Under simple random validation, sampling weights were set to one. Bias, empirical 95\% coverage, and average interval width were calculated on the log-HR scale. Fit failures, validation-pattern frequencies, effective sample sizes, and weight distributions were retained as diagnostics. The common time basis was constructed from validated and CP-derived person-period times, so it could change slightly across paired noise levels.

\paragraph{Censoring and covariance conventions.}
Subjects contributed person-period rows through the interval containing their event or censoring time. Treating a partially observed terminal censoring interval as a full event-free interval introduces a coarsening approximation. Standard errors used patient-level estimating-function contributions rather than treating person-period rows as independent. The covariance calculation is conditional on the fitted CP and validation models, as described in Sections~\ref{supp:s5}--\ref{supp:s8}.

\clearpage
\section*{Supplementary Results}
\subsection{Chart-Reviewed Cohort Characteristics}
\label{supp:s10}
Supplementary Table~\ref{tab:s1-chart-reviewed-cohort} describes the 1026-patient Northwestern Medicine ICI cohort, overall and by chart-reviewed irAE status. Clinicians recorded adjudicated labels in REDCap \citep{harris2009redcap}. Validated survival outcomes and autoimmune disease (AID) history were available for all patients before masking.

\begin{table}[!htbp]
\caption{Characteristics of the chart-reviewed ICI cohort, overall and by chart-reviewed irAE status}
\label{tab:s1-chart-reviewed-cohort}
\centering\small
\setlength{\tabcolsep}{4pt}
\renewcommand{\arraystretch}{1.13}
\begin{tabular*}{\textwidth}{@{\extracolsep{\fill}}p{0.39\textwidth}rrr@{}}
\toprule
Characteristic & Overall & \shortstack[r]{No qualifying\\irAE} & \shortstack[r]{Qualifying\\irAE} \\
 & $N=1026$ & $n=522$ & $n=504$ \\
\midrule
\addlinespace[0.55em]
\multicolumn{4}{@{}l}{\textbf{Cohort}} \\
Patients, n & 1026 & 522 & 504 \\
\addlinespace[0.55em]
\multicolumn{4}{@{}l}{\textbf{Age at ICI initiation, years}} \\
Mean (SD) & 66.9 (12.4) & 67.7 (12.1) & 66.1 (12.6) \\
Median [IQR] & 68.2 [60.2, 75.4] & 69.2 [61.4, 76.0] & 67.2 [58.3, 74.9] \\
\addlinespace[0.55em]
\multicolumn{4}{@{}l}{\textbf{Sex/gender, n (\%)}} \\
Male & 541 (52.7) & 285 (54.6) & 256 (50.8) \\
Female & 485 (47.3) & 237 (45.4) & 248 (49.2) \\
\addlinespace[0.55em]
\multicolumn{4}{@{}l}{\textbf{Initial ICI regimen, n (\%)}} \\
PD-1 only & 668 (65.1) & 365 (69.9) & 303 (60.1) \\
PD-L1 only & 180 (17.5) & 98 (18.8) & 82 (16.3) \\
CTLA-4-containing or combination & 178 (17.3) & 59 (11.3) & 119 (23.6) \\
\addlinespace[0.55em]
\multicolumn{4}{@{}l}{\textbf{Pre-existing AID, n (\%)}} \\
No & 946 (92.2) & 481 (92.1) & 465 (92.3) \\
Yes & 80 (7.8) & 41 (7.9) & 39 (7.7) \\
\addlinespace[0.55em]
\multicolumn{4}{@{}l}{\textbf{Chart-reviewed irAE occurrence, n (\%)}} \\
Event & 504 (49.1) & 0 (0.0) & 504 (100.0) \\
No event during observed follow-up & 522 (50.9) & 522 (100.0) & 0 (0.0) \\
\addlinespace[0.55em]
\multicolumn{4}{@{}l}{\textbf{Observation time, days}} \\
Available follow-up, median [IQR] & 386.0 [147.2, 730.0] & 277.0 [92.0, 679.8] & 496.0 [246.0, 730.0] \\
Time to event or censoring, median [IQR] & 119.5 [42.0, 353.0] & 277.0 [92.0, 679.8] & 67.0 [25.0, 147.0] \\
Time to event among cases, median [IQR] & 67.0 [25.0, 147.0] & Not applicable & 67.0 [25.0, 147.0] \\
\bottomrule
\end{tabular*}
\par\vspace{0.8em}
\begin{minipage}{\textwidth}
\footnotesize\raggedright
Values are n (\%) or the indicated summary; percentages use column denominators. AID, autoimmune disease; SD, standard deviation; IQR, 25th-75th percentiles; ICI, immune checkpoint inhibitor; irAE, immune-related adverse event.\par\smallskip
The full unmasked analytic cohort has chart-reviewed outcome and AID-history labels, CP outputs, positive outcome times, and complete model covariates. No displayed variable is missing. A qualifying event is an irAE in a clinical category not documented before ICI initiation. Patients without a qualifying event may have other irAEs during follow-up.\par\smallskip
Follow-up is measured from first ICI initiation to the earlier of last recorded follow-up and two years. Analysis time ends at the qualifying event or censoring. Time to event is summarized only among cases; 'Not applicable' is not missing data.\par\smallskip
\end{minipage}
\end{table}

\clearpage
\subsection{Clinical CPs and Masked-Validation Analyses}
\label{supp:s11}
The outcome was time from first ICI treatment to the first irAE category not documented before treatment. Follow-up ended at the earlier of the last recorded follow-up and two years. Outcome CPs used post-treatment diagnosis codes and the same adjudicated baseline-category history used to define the validated outcome. The AID CP required at least one AID ICD-10 diagnosis in the two-year pretreatment window. These CPs were fixed before labels were masked; the experiment did not retrain the CP algorithms.

Clinical survival models used 30-day intervals, a 3-degree-of-freedom natural-spline baseline time effect, and covariates for pre-existing AID, ICI class, age, and sex. PD-1-only therapy was the reference regimen. The outcome CP had sensitivity 85.1\%, specificity 33.3\%, positive predictive value 55.2\%, and negative predictive value 69.9\%. The corresponding values for the AID CP were 26.3\%, 97.8\%, 50.0\%, and 94.0\%. These measures assess binary labels and do not establish event-time accuracy.

For the EHR-guided masking experiment, outcome-label masking targeted 20\%, 30\%, 50\%, or 70\% and AID-label masking targeted 20\%, 30\%, or 50\%. Logistic masking probabilities depended on age, sex, and ICI class; the log odds of masking increased by $\log(1.5)$ per standard deviation of the prespecified masking score. Each combination used 100 masked-validation replicates. Draws failing minimum support requirements were resampled, with up to 50 attempts. Each component dataset required at least 3 AID-positive patients, 10 AID-negative patients, 5 events, and 5 nonevents; fits additionally required at least 20 patients and 20 person-period records. Estimated validation probabilities were floored at 0.01 before constructing stabilized inverse-probability weights. Covariance estimation used patients as clusters and treated the estimated validation weights as fixed.

\begin{figure}[p]
\centering
\includegraphics[width=\linewidth]{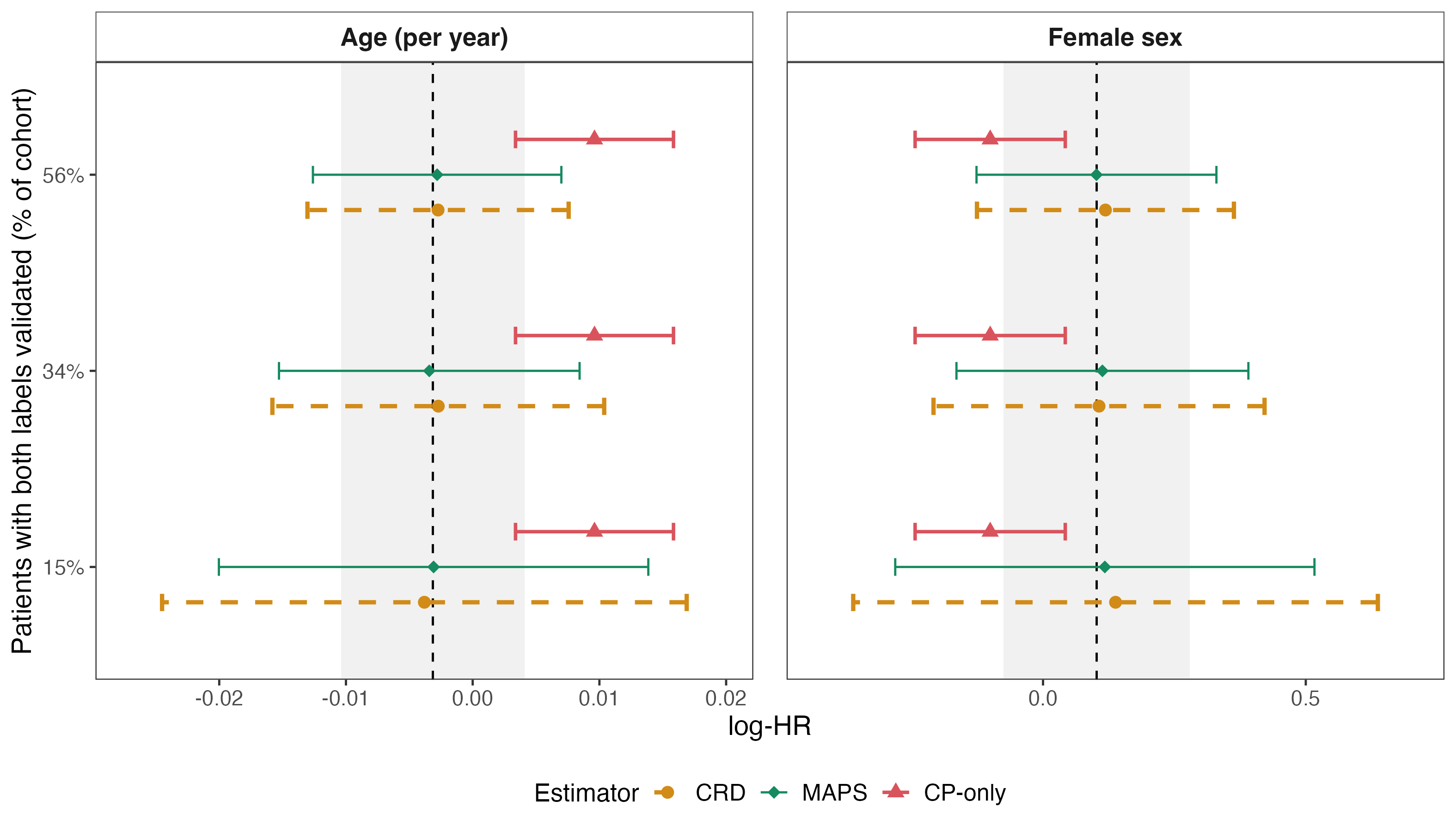}
\caption{Masked-validation estimates for age and sex under EHR-guided validation. Points and intervals show mean log-HRs and mean lower and upper 95\% confidence interval limits across 100 masking replicates per setting. Approximately 56\%, 34\%, and 15\% of the 1026-patient cohort had both outcome and AID-history labels validated. Age is expressed per year, and male sex is the reference category. The dashed line and gray band show the fully validated estimate and its 95\% confidence interval.}
\label{fig:s-ici-remaining-forest}
\end{figure}

\begin{figure}[p]
\centering
\includegraphics[width=\linewidth]{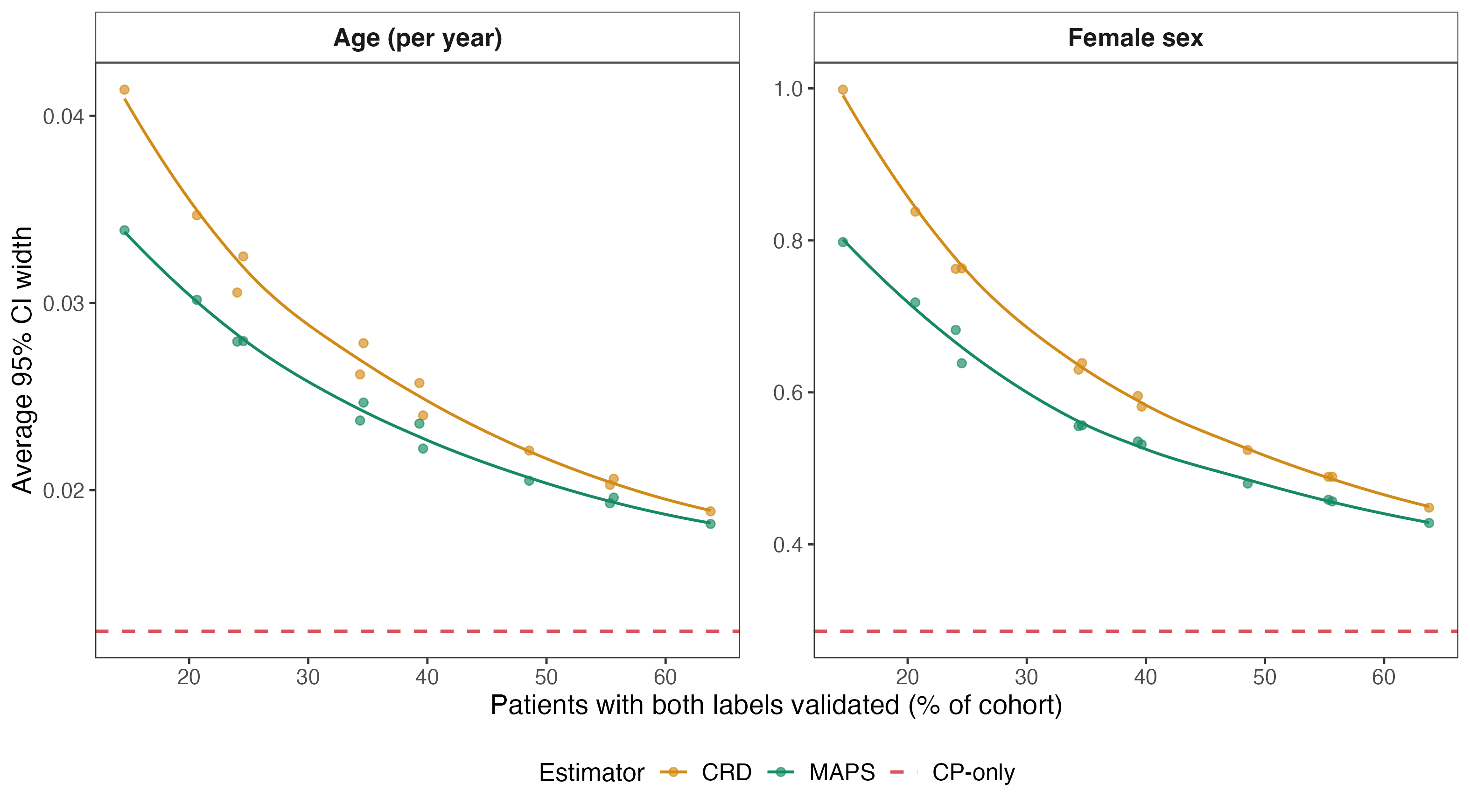}
\caption{Average 95\% confidence interval width for age and sex under EHR-guided validation. Points show mean interval widths on the log-HR scale across 100 masking replicates per configuration. The horizontal axis gives the mean percentage of the 1026-patient cohort with both outcome and AID-history labels validated. Curves are smoothed guides; dashed lines show CP-only interval widths.}
\label{fig:s-ici-remaining-width}
\end{figure}

Supplementary Table~\ref{tab:s2-clinical-reference} gives the fully validated and CP-only estimates for all model covariates. Supplementary Table~\ref{tab:s3-clinical-masked} gives CRD and MAPS estimates for the three representative settings in the main clinical figures. Figures~\ref{fig:s-ici-remaining-forest} and~\ref{fig:s-ici-remaining-width} show the corresponding estimates and interval widths for age and sex, completing the covariates shown in the main figures. Age coefficients are expressed per year, and male sex is the reference category. The fully validated analysis is an empirical benchmark, not a known true clinical effect.

\begin{table}[!htbp]
\centering\small
\caption{Fully validated and CP-only estimates for all clinical model covariates.}
\label{tab:s2-clinical-reference}
\begin{tabular}{@{}p{0.40\linewidth}rr@{}}
\toprule
Covariate & Fully validated HR (95\% CI) & CP-only HR (95\% CI) \\
\midrule
Pre-existing AID & 0.93 (0.67--1.29) & 1.53 (1.12--2.09) \\
CTLA-4-containing or combination regimen & 1.91 (1.53--2.38) & 1.41 (1.15--1.72) \\
PD-L1-only regimen & 0.99 (0.77--1.26) & 0.98 (0.81--1.19) \\
Age, per year & 0.997 (0.990--1.004) & 1.010 (1.003--1.016) \\
Female sex & 1.11 (0.93--1.32) & 0.90 (0.78--1.04) \\
\bottomrule
\end{tabular}
\par\smallskip
\begin{minipage}{\linewidth}\footnotesize
AID, autoimmune disease; HR, hazard ratio; CI, confidence interval. PD-1-only therapy and male sex are the reference categories. Both analyses use the full cohort of 1026 patients. CP-only analysis substitutes CP-derived outcome and AID-history labels for their validated counterparts.
\end{minipage}
\end{table}

\clearpage
\begin{table}[!htbp]
\centering\small
\caption{Masked-validation estimates for all clinical covariates at selected validation levels.}
\label{tab:s3-clinical-masked}
\begin{tabular}{@{}p{0.38\linewidth}rr@{}}
\toprule
Covariate & CRD log-HR (95\% CI) & MAPS log-HR (95\% CI) \\
\midrule
\multicolumn{3}{@{}l}{\textbf{Approximately 56\% fully validated (mean $n=571$)}} \\
Pre-existing AID & -0.04 (-0.50--0.43) & -0.05 (-0.50--0.41) \\
CTLA-4-containing or combination regimen & 0.67 (0.35--0.98) & 0.67 (0.38--0.96) \\
PD-L1-only regimen & -0.01 (-0.33--0.32) & -0.02 (-0.33--0.28) \\
Age, per year & -0.003 (-0.013--0.008) & -0.003 (-0.013--0.007) \\
Female sex & 0.12 (-0.13--0.36) & 0.10 (-0.13--0.33) \\
\addlinespace[0.5em]
\multicolumn{3}{@{}l}{\textbf{Approximately 34\% fully validated (mean $n=352$)}} \\
Pre-existing AID & -0.09 (-0.70--0.52) & -0.13 (-0.72--0.46) \\
CTLA-4-containing or combination regimen & 0.67 (0.25--1.09) & 0.66 (0.30--1.02) \\
PD-L1-only regimen & -0.03 (-0.45--0.40) & -0.04 (-0.43--0.34) \\
Age, per year & -0.003 (-0.016--0.010) & -0.003 (-0.015--0.008) \\
Female sex & 0.11 (-0.21--0.42) & 0.11 (-0.16--0.39) \\
\addlinespace[0.5em]
\multicolumn{3}{@{}l}{\textbf{Approximately 15\% fully validated (mean $n=149$)}} \\
Pre-existing AID & -0.06 (-0.99--0.88) & -0.04 (-0.92--0.84) \\
CTLA-4-containing or combination regimen & 0.69 (0.03--1.34) & 0.69 (0.19--1.20) \\
PD-L1-only regimen & 0.01 (-0.66--0.67) & 0.00 (-0.55--0.55) \\
Age, per year & -0.004 (-0.025--0.017) & -0.003 (-0.020--0.014) \\
Female sex & 0.14 (-0.36--0.64) & 0.12 (-0.28--0.52) \\
\addlinespace[0.5em]
\bottomrule
\end{tabular}
\par\smallskip
\begin{minipage}{\linewidth}\footnotesize
AID, autoimmune disease; CRD, chart-reviewed data. Each entry gives the mean log-HR estimate and the mean lower and upper 95\% confidence interval limits across 100 masking replicates; these are not confidence intervals for the Monte Carlo mean. Outcome/AID-label masking targeted 20\%/30\%, 50\%/30\%, and 70\%/50\%, respectively. Validation percentages and sample sizes are rounded averages across replicates. PD-1-only therapy and male sex are the reference categories.
\end{minipage}
\end{table}

\clearpage

\clearpage
\subsection{Random Validation of Survival Outcomes Only}
\label{supp:s12}
We also evaluated MAPS when only survival outcomes were masked and all covariates remained validated. This analysis used the same 1026-patient cohort, outcome CP, covariates, and 30-day intervals as the main clinical analysis. Within chart-reviewed event-status strata, we randomly masked 20\%, 30\%, 50\%, 70\%, 80\%, or 90\% of outcome labels, with 100 replicates per configuration. All 600 replicates completed successfully. Sampling weights were one, so CRD reduced to ordinary complete-case analysis. Patient-level sandwich estimation retained covariance between estimates from the same labeled patients and set cross-covariances with the disjoint unlabeled group to zero.

\begin{figure}[p]
\centering
\includegraphics[width=\linewidth]{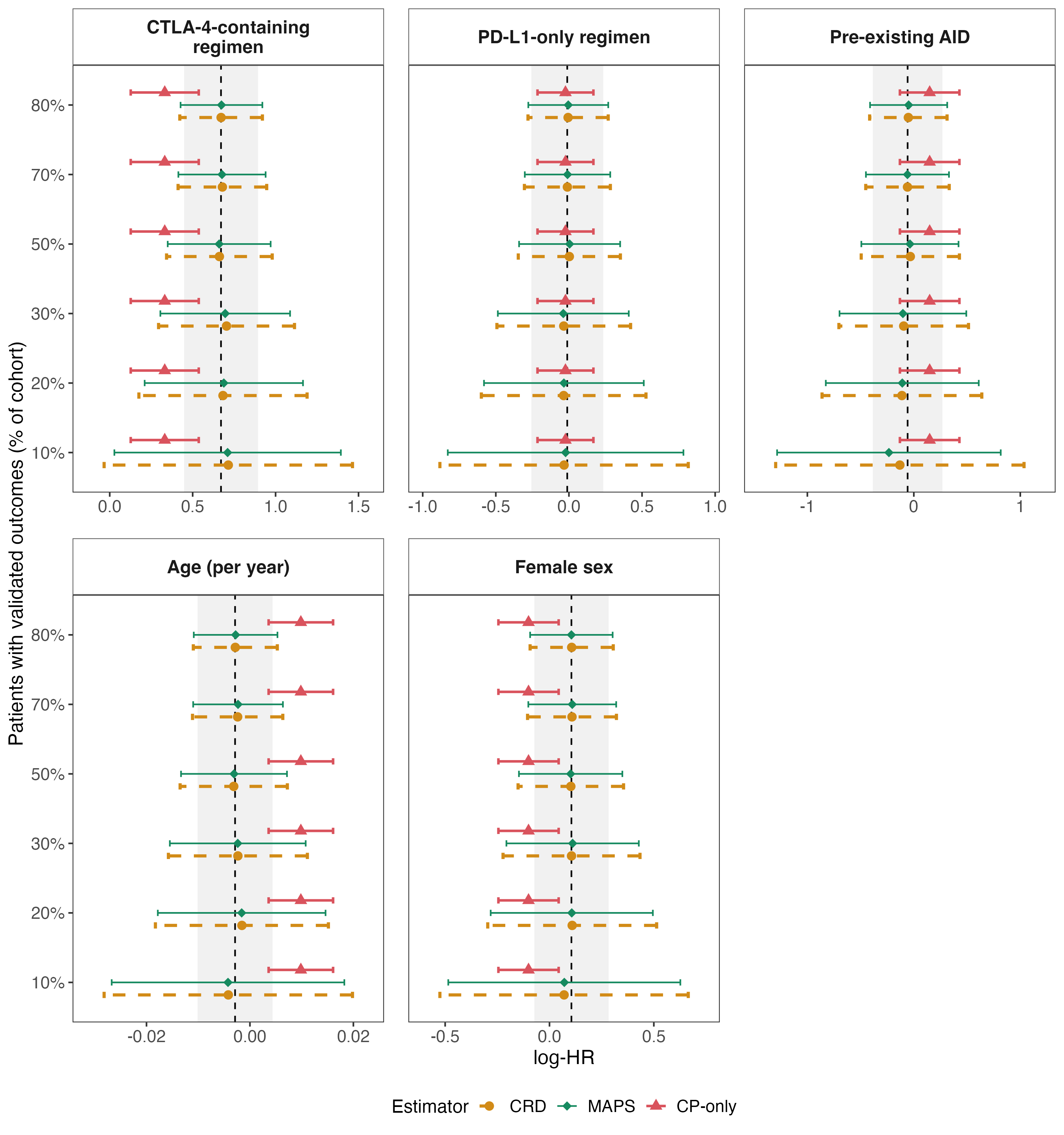}
\caption{Clinical estimates with randomly masked survival outcomes and fully validated covariates. Points and intervals show mean log-HRs and mean lower and upper 95\% confidence interval limits across 100 masking replicates per configuration. Vertical-axis labels give the rounded percentage of the 1026-patient cohort with validated outcomes. Reference categories are PD-1-only therapy, no pre-existing AID, and male sex; age is expressed per year. Dashed lines and gray bands show fully validated estimates and 95\% confidence intervals.}
\label{fig:s-outcome-only-forest}
\end{figure}

\begin{figure}[p]
\centering
\includegraphics[width=\linewidth]{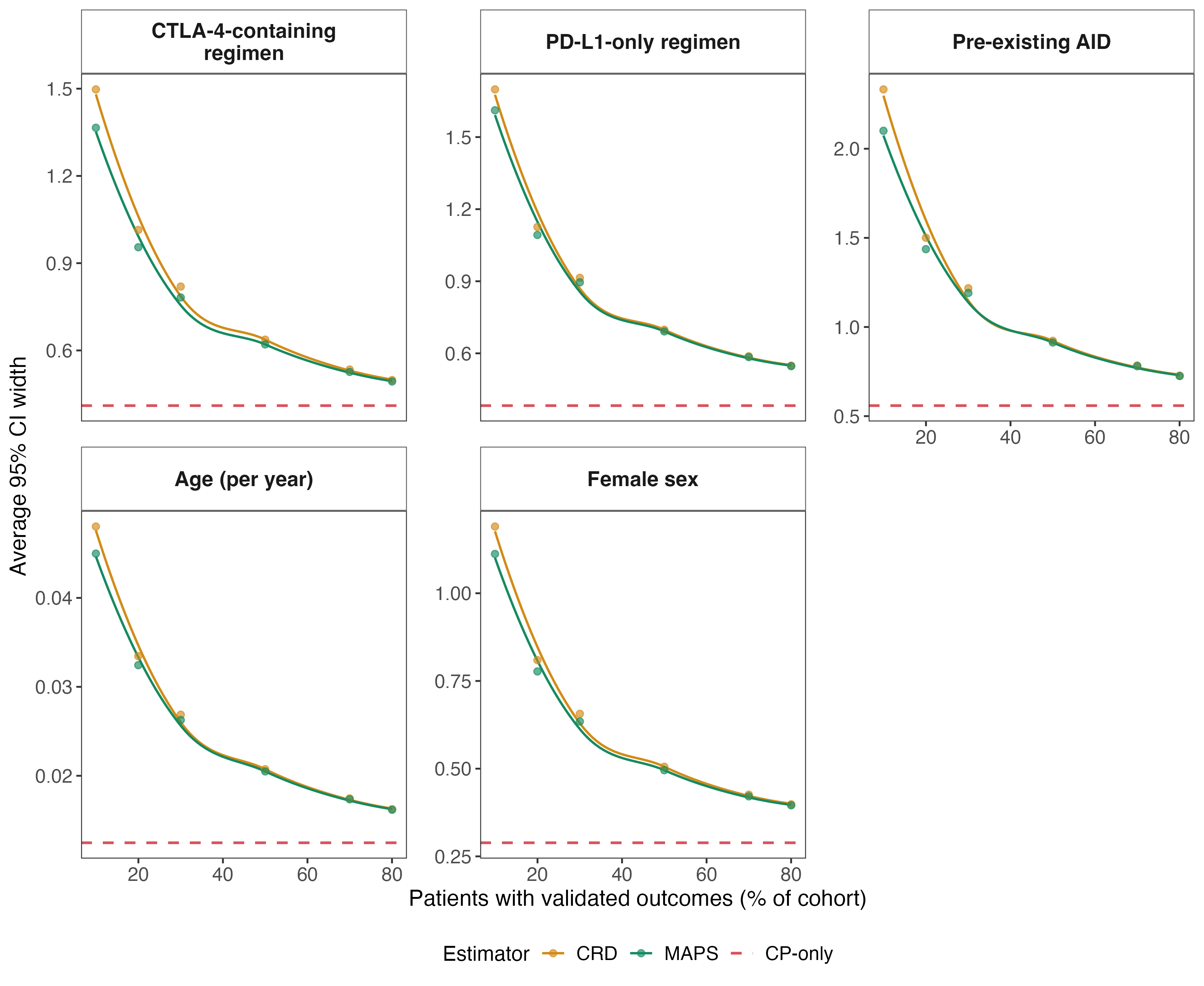}
\caption{Average 95\% confidence interval width with randomly masked survival outcomes and fully validated covariates. Points show mean widths on the log-HR scale across 100 masking replicates per configuration. The horizontal axis gives the mean percentage of the 1026-patient cohort with validated outcomes; all covariates remained validated. Curves are smoothed guides; dashed lines show CP-only interval widths.}
\label{fig:s-outcome-only-width}
\end{figure}

Figures~\ref{fig:s-outcome-only-forest} and~\ref{fig:s-outcome-only-width} show all model coefficients and interval widths. The AID covariate was validated in every patient, so CP-only analysis replaced only the survival outcome. Each analysis used its own fully validated estimate and confidence interval as the benchmark. MAPS intervals were narrower than CRD intervals, with reductions in mean width of 0.3\%--10.0\% across the five coefficients and evaluated validation levels. These outcome-only experiments are distinct from the main analysis, in which both outcomes and AID-history labels could be unavailable.

\clearpage
\subsection{Alternative AID CPs Under Random Validation}
\label{supp:s13}
We examined two AID CPs in the same cohort. The single-code rule required at least one AID ICD-10 diagnosis date during the two years before ICI initiation. The repeated-code rule required at least two distinct diagnosis dates at least 30 days apart within the same window. The outcome CP was unchanged. Table~\ref{tab:s4-autoimmune-cps} compares both rules with chart-reviewed AID history. The repeated-code rule increased specificity but reduced sensitivity.

Outcome and AID-history labels were randomly masked within strata defined jointly by chart-reviewed event status and AID history, using unit sampling weights and 100 replicates per configuration. For the single-code CP, outcome masking targeted 10\%, 30\%, 50\%, or 70\% and AID-history masking targeted 10\%, 20\%, or 30\%. For the repeated-code CP, the corresponding grids were 30\%, 50\%, or 70\% and 20\% or 30\%. Models used the same covariates and 30-day intervals as the main analysis, with a 3-degree-of-freedom natural-spline baseline time effect. Masks were redrawn up to 50 times to meet the component-dataset support requirements in Section~\ref{supp:s11}; replicates that still lacked support or failed fitting were retained as failures.

\begin{figure}[p]
\centering
\includegraphics[width=\linewidth]{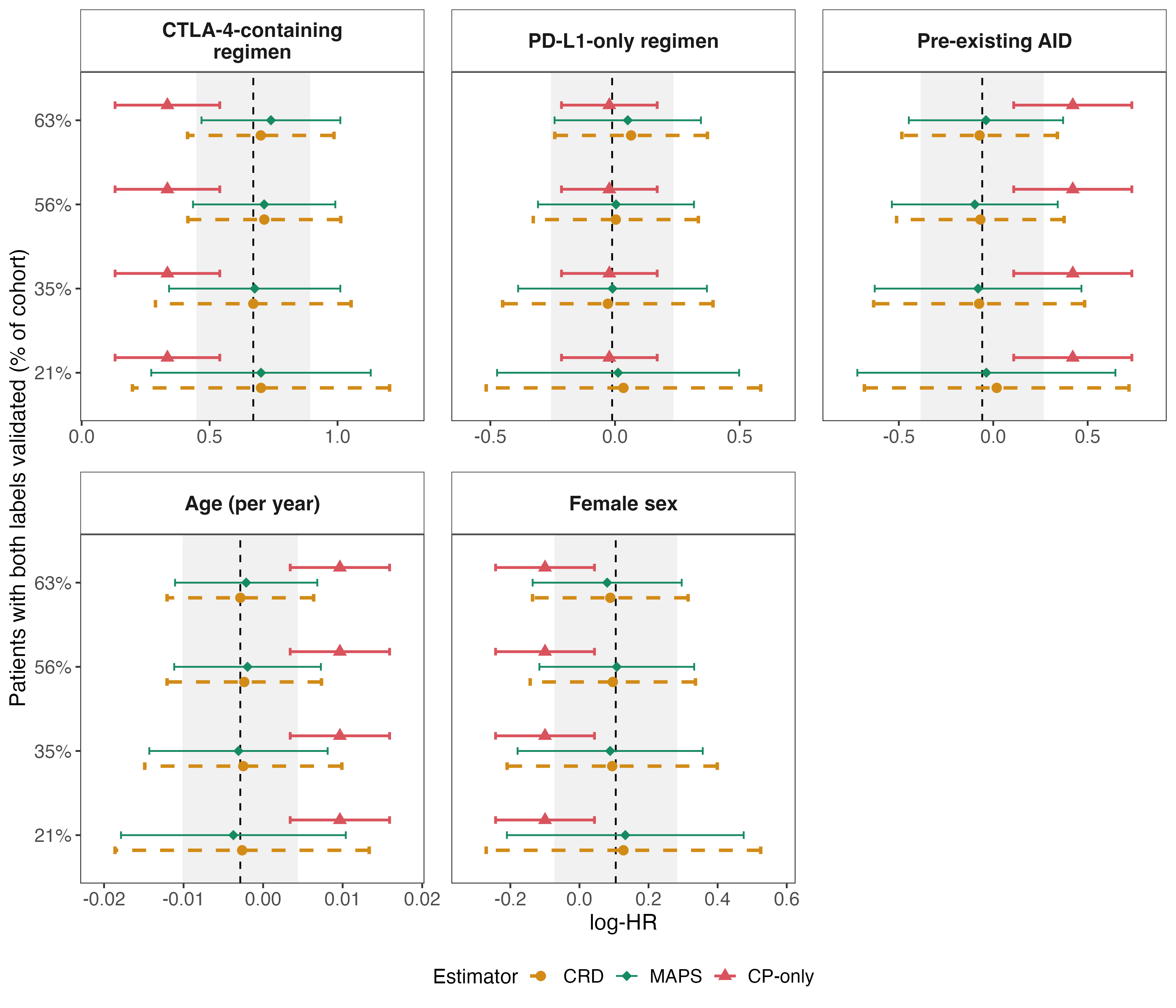}
\caption{Clinical estimates under random validation using the single-code AID CP. Points and intervals show mean log-HRs and mean lower and upper 95\% confidence interval limits from successful fits; rounded percentages indicate the proportion of the 1026-patient cohort with both outcome and AID-history labels validated. Reference categories are PD-1-only therapy, no pre-existing AID, and male sex; age is expressed per year. Dashed lines and gray bands show fully validated estimates and 95\% confidence intervals. Successful-fit counts are given in Table~\ref{tab:s5-mcar-support}.}
\label{fig:s-any-forest}
\end{figure}

\begin{figure}[p]
\centering
\includegraphics[width=\linewidth]{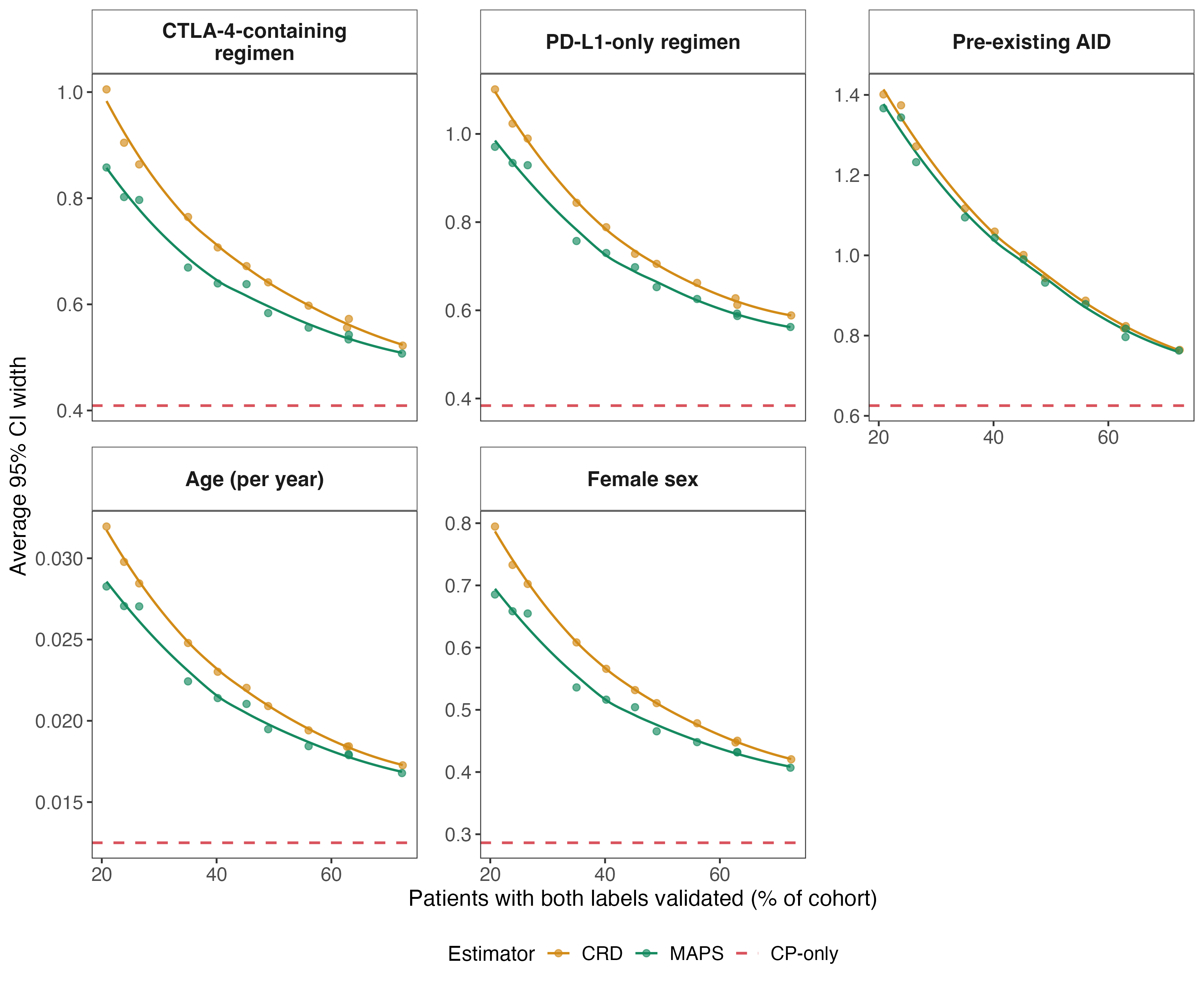}
\caption{Average 95\% confidence interval width under random validation using the single-code AID CP. Points show mean interval widths on the log-HR scale from successful fits. The horizontal axis gives the mean percentage of the 1026-patient cohort with both outcome and AID-history labels validated in each configuration. Curves are smoothed guides; dashed lines show CP-only interval widths.}
\label{fig:s-any-width}
\end{figure}

\begin{figure}[p]
\centering
\includegraphics[width=\linewidth]{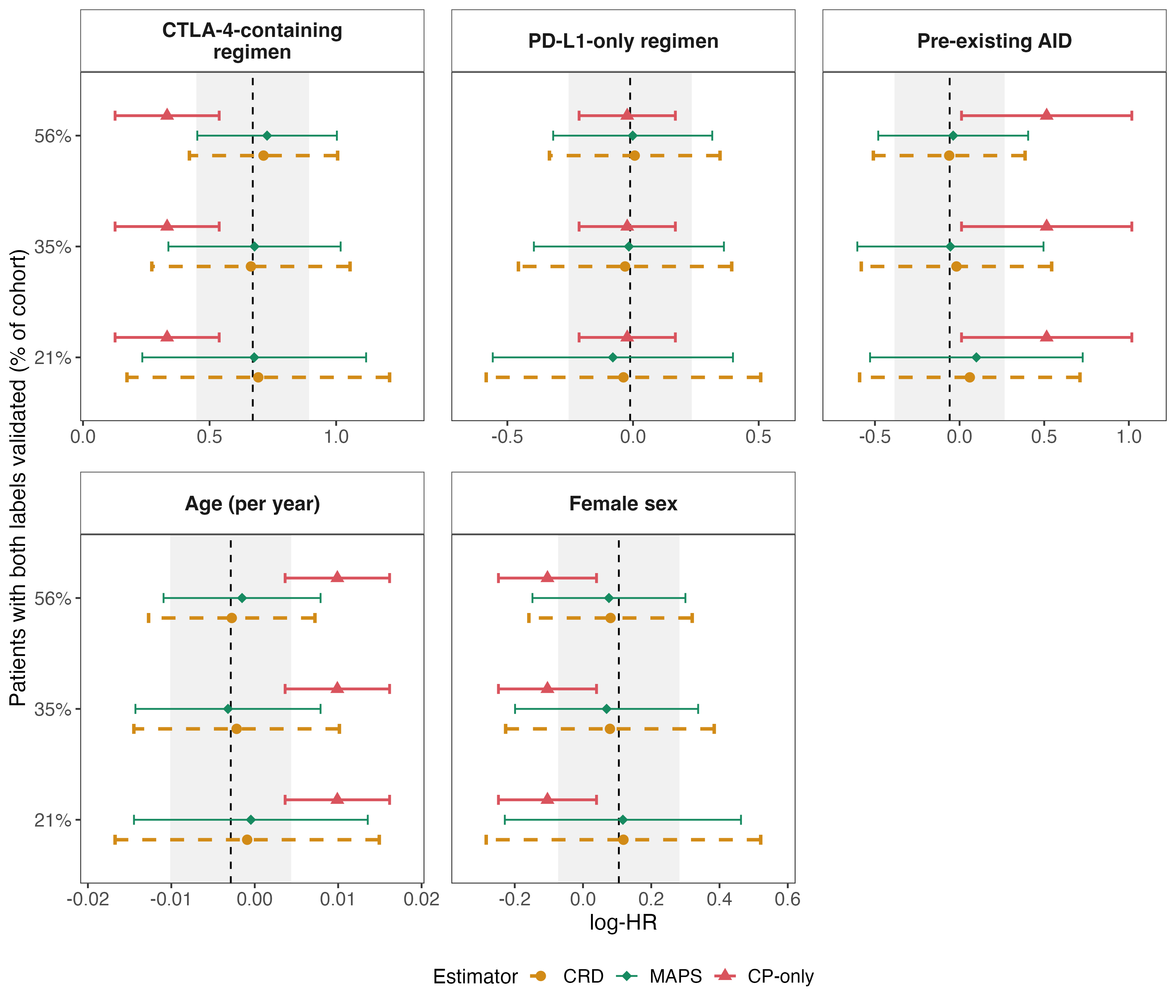}
\caption{Clinical estimates under random validation using the repeated-code AID CP, requiring diagnosis dates at least 30 days apart. Points and intervals show mean log-HRs and mean lower and upper 95\% confidence interval limits from successful fits; rounded percentages indicate the proportion of the 1026-patient cohort with both outcome and AID-history labels validated. Reference categories are PD-1-only therapy, no pre-existing AID, and male sex; age is expressed per year. Dashed lines and gray bands show fully validated estimates and 95\% confidence intervals. Successful-fit counts are given in Table~\ref{tab:s5-mcar-support}.}
\label{fig:s-repeat-forest}
\end{figure}

\begin{figure}[p]
\centering
\includegraphics[width=\linewidth]{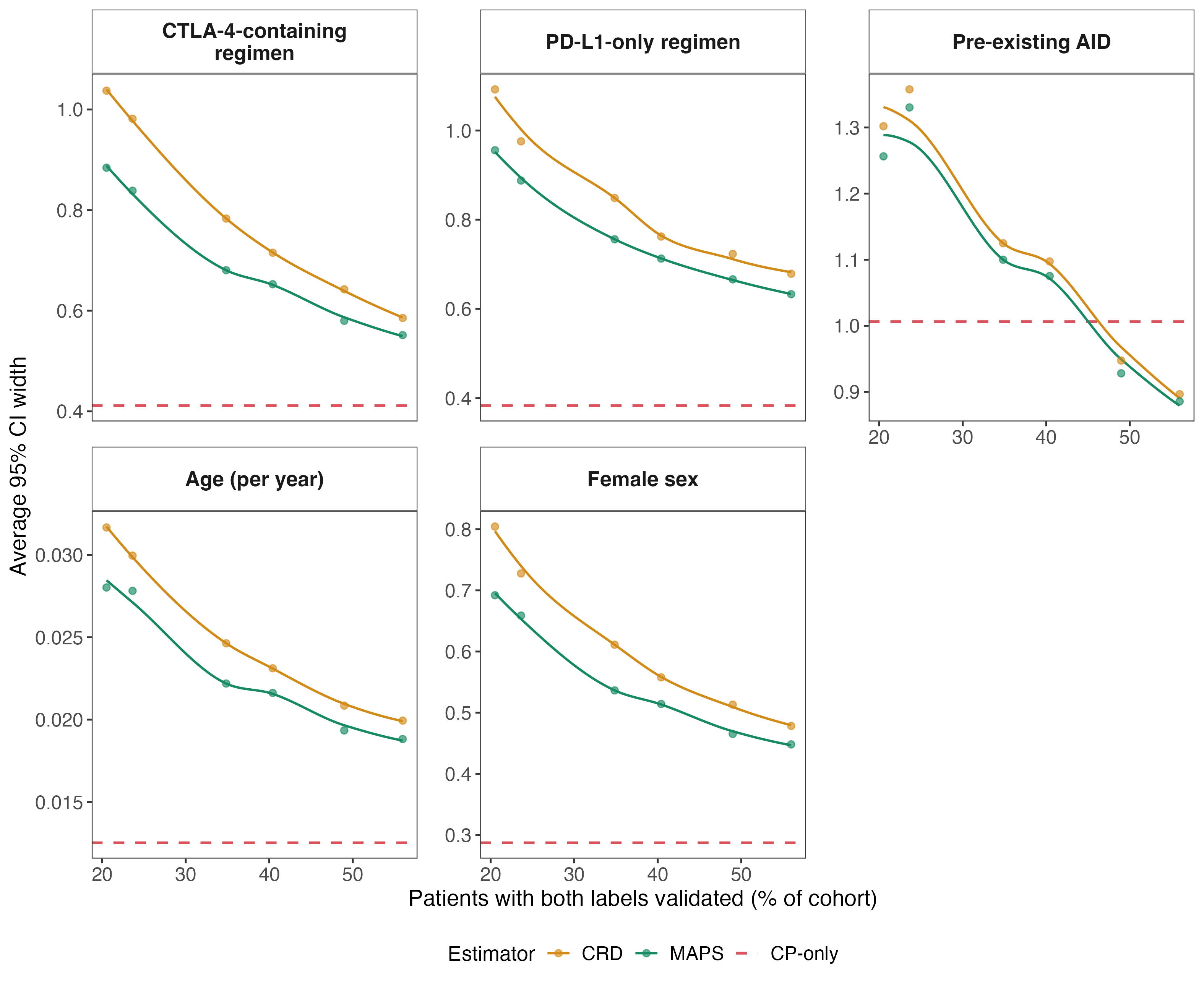}
\caption{Average 95\% confidence interval width under random validation using the repeated-code AID CP. Points show mean interval widths on the log-HR scale from successful fits. The horizontal axis gives the mean percentage of the 1026-patient cohort with both outcome and AID-history labels validated in each configuration. Curves are smoothed guides; dashed lines show CP-only interval widths.}
\label{fig:s-repeat-width}
\end{figure}

Figures~\ref{fig:s-any-forest}--\ref{fig:s-repeat-width} show all model covariates, using the fully validated fit from each saved analysis as its benchmark. Estimates and interval widths summarize successful fits for each method. Precision gains for AID history were modest under both CP rules, with mean width reductions of 0.2\%--3.5\% relative to CRD. Some MAPS fits were unavailable because of sparse support or nonconvergence; Table~\ref{tab:s5-mcar-support} reports successful-fit counts for every configuration. Configurations without any successful MAPS fit are omitted from the plots. The saved MCAR implementation used patient-level sandwich covariances, retaining covariance among fits from the same fully validated patients and setting cross-covariances between disjoint patient groups to zero. These calculations were conditional on the fixed CPs; the main EHR-guided analysis used the joint covariance calculation in Section~\ref{supp:s7}.

\begin{table}[!htbp]
\caption{AID CP performance against chart-reviewed history in 1026 patients}
\label{tab:s4-autoimmune-cps}
\centering\small
\setlength{\tabcolsep}{4pt}
\renewcommand{\arraystretch}{1.15}
\begin{tabular*}{\textwidth}{@{\extracolsep{\fill}}lrrrrr@{}}
\toprule
CP rule & Positive, n (\%) & Sensitivity & Specificity & PPV & NPV \\
\midrule
Single-code & 42 (4.1) & 26.3\% & 97.8\% & 50.0\% & 94.0\% \\
Repeated-code & 21 (2.0) & 17.5\% & 99.3\% & 66.7\% & 93.4\% \\
\bottomrule
\end{tabular*}
\par\vspace{0.7em}
\begin{minipage}{\textwidth}\footnotesize\raggedright
AID, autoimmune disease. Both rules use AID ICD-10 diagnosis dates during the two years before ICI initiation. The single-code rule requires at least one date; the repeated-code rule requires at least two distinct dates separated by at least 30 days. Chart review identified 80 patients with pre-existing AID. PPV, positive predictive value; NPV, negative predictive value.
\end{minipage}
\end{table}

\clearpage
\begin{table}[!htbp]
\caption{Validation availability and successful MAPS fits in the four-pattern random-validation analyses}
\label{tab:s5-mcar-support}
\centering\small
\setlength{\tabcolsep}{4pt}
\renewcommand{\arraystretch}{1.12}
\begin{tabular*}{\textwidth}{@{\extracolsep{\fill}}lrrrr@{}}
\toprule
CP rule & \shortstack{Outcome\\masked (\%)} & \shortstack{AID history\\masked (\%)} & \shortstack{Mean fully\\validated n} & \shortstack{MAPS fits\\of 100} \\
\midrule
Single-code & 10 & 10 & 835.5 & 0 \\
Single-code & 10 & 20 & 743.1 & 65 \\
Single-code & 10 & 30 & 646.8 & 100 \\
Single-code & 30 & 10 & 643.6 & 67 \\
Single-code & 30 & 20 & 574.9 & 100 \\
Single-code & 30 & 30 & 502.6 & 100 \\
Single-code & 50 & 10 & 463.9 & 100 \\
Single-code & 50 & 20 & 412.4 & 100 \\
Single-code & 50 & 30 & 359.4 & 100 \\
Single-code & 70 & 10 & 272.1 & 100 \\
Single-code & 70 & 20 & 245.0 & 100 \\
Single-code & 70 & 30 & 213.6 & 100 \\
\addlinespace
Repeated-code & 30 & 20 & 574.3 & 88 \\
Repeated-code & 30 & 30 & 502.5 & 100 \\
Repeated-code & 50 & 20 & 414.6 & 100 \\
Repeated-code & 50 & 30 & 357.5 & 100 \\
Repeated-code & 70 & 20 & 242.5 & 100 \\
Repeated-code & 70 & 30 & 210.5 & 100 \\
\bottomrule
\end{tabular*}
\par\vspace{0.7em}
\begin{minipage}{\textwidth}\footnotesize\raggedright
AID, autoimmune disease; CRD, chart-reviewed data. Each configuration included 100 masking replicates from the 1026-patient cohort. CRD completed all 100 fits in every configuration. Mean fully validated counts use all attempted replicates. MAPS failures arose from insufficient component-dataset support or fitting failure; no estimate was substituted for a failed fit. Plotted summaries use each method's successful fits, so the contributing replicate sets can differ.
\end{minipage}
\end{table}

\clearpage

\clearpage

\clearpage
\section*{Additional Simulation Results}

\clearpage

\clearpage
\begin{table}[!htbp]
\caption{Standard-error calibration in the outcome prediction-noise analysis}
\label{tab:s6-simulation-calibration}
\centering\small
\setlength{\tabcolsep}{3pt}
\renewcommand{\arraystretch}{1.15}
\begin{tabular*}{\textwidth}{@{\extracolsep{\fill}}llrrrrrr@{}}
\toprule
 & & \multicolumn{3}{c}{MAPS} & \multicolumn{3}{c}{CRD} \\
\cmidrule(lr){3-5}\cmidrule(lr){6-8}
$\sigma$ & Coefficient & \shortstack{Empirical\\SD} & \shortstack{Mean\\SE} & \shortstack{Coverage\\(MCSE)} & \shortstack{Empirical\\SD} & \shortstack{Mean\\SE} & \shortstack{Coverage\\(MCSE)} \\
\midrule
0.00 & $\beta_{1}$ & 0.032 & 0.028 & 91.3 (1.6) & 0.079 & 0.077 & 95.3 (1.2) \\
0.00 & $\beta_{2}$ & 0.031 & 0.028 & 91.0 (1.7) & 0.079 & 0.078 & 94.7 (1.3) \\
\addlinespace[2pt]
0.10 & $\beta_{1}$ & 0.031 & 0.029 & 92.3 (1.5) & 0.079 & 0.077 & 95.3 (1.2) \\
0.10 & $\beta_{2}$ & 0.032 & 0.029 & 93.0 (1.5) & 0.079 & 0.078 & 94.7 (1.3) \\
\addlinespace[2pt]
0.20 & $\beta_{1}$ & 0.033 & 0.032 & 93.0 (1.5) & 0.079 & 0.077 & 95.3 (1.2) \\
0.20 & $\beta_{2}$ & 0.034 & 0.032 & 93.0 (1.5) & 0.079 & 0.078 & 94.7 (1.3) \\
\addlinespace[2pt]
0.30 & $\beta_{1}$ & 0.037 & 0.036 & 94.0 (1.4) & 0.079 & 0.077 & 95.3 (1.2) \\
0.30 & $\beta_{2}$ & 0.037 & 0.035 & 94.3 (1.3) & 0.079 & 0.078 & 94.7 (1.3) \\
\addlinespace[2pt]
0.50 & $\beta_{1}$ & 0.046 & 0.044 & 94.7 (1.3) & 0.079 & 0.077 & 95.3 (1.2) \\
0.50 & $\beta_{2}$ & 0.047 & 0.044 & 94.0 (1.4) & 0.079 & 0.078 & 94.7 (1.3) \\
\addlinespace[2pt]
0.75 & $\beta_{1}$ & 0.055 & 0.052 & 95.0 (1.3) & 0.079 & 0.077 & 95.3 (1.2) \\
0.75 & $\beta_{2}$ & 0.055 & 0.052 & 91.7 (1.6) & 0.079 & 0.078 & 94.7 (1.3) \\
\addlinespace[2pt]
1.00 & $\beta_{1}$ & 0.062 & 0.057 & 92.3 (1.5) & 0.079 & 0.077 & 95.3 (1.2) \\
1.00 & $\beta_{2}$ & 0.059 & 0.057 & 93.3 (1.4) & 0.079 & 0.078 & 94.7 (1.3) \\
\bottomrule
\end{tabular*}
\par\vspace{0.7em}
\begin{minipage}{\textwidth}\footnotesize\raggedright
CRD, chart-reviewed data. Each noise level used 300 successful replicates with $N=5000$, 10\% outcome validation, and $\varepsilon_i\sim N(0.30,\sigma^2)$. The true coefficients were $\beta_1=\log(1.4)$ and $\beta_2=\log(0.75)$. Empirical SD is the standard deviation of coefficient estimates across replicates; mean SE averages the estimated standard errors. Coverage and its Monte Carlo standard error (MCSE) are percentages, with $\operatorname{MCSE}=100\sqrt{\hat c(1-\hat c)/300}$ for coverage proportion $\hat c$. All SDs and SEs are on the log-HR scale. The CSV companion retains full saved precision and includes mean bias and its MCSE.
\end{minipage}
\end{table}

\begin{figure}[!htbp]
\centering
\includegraphics[width=\linewidth]{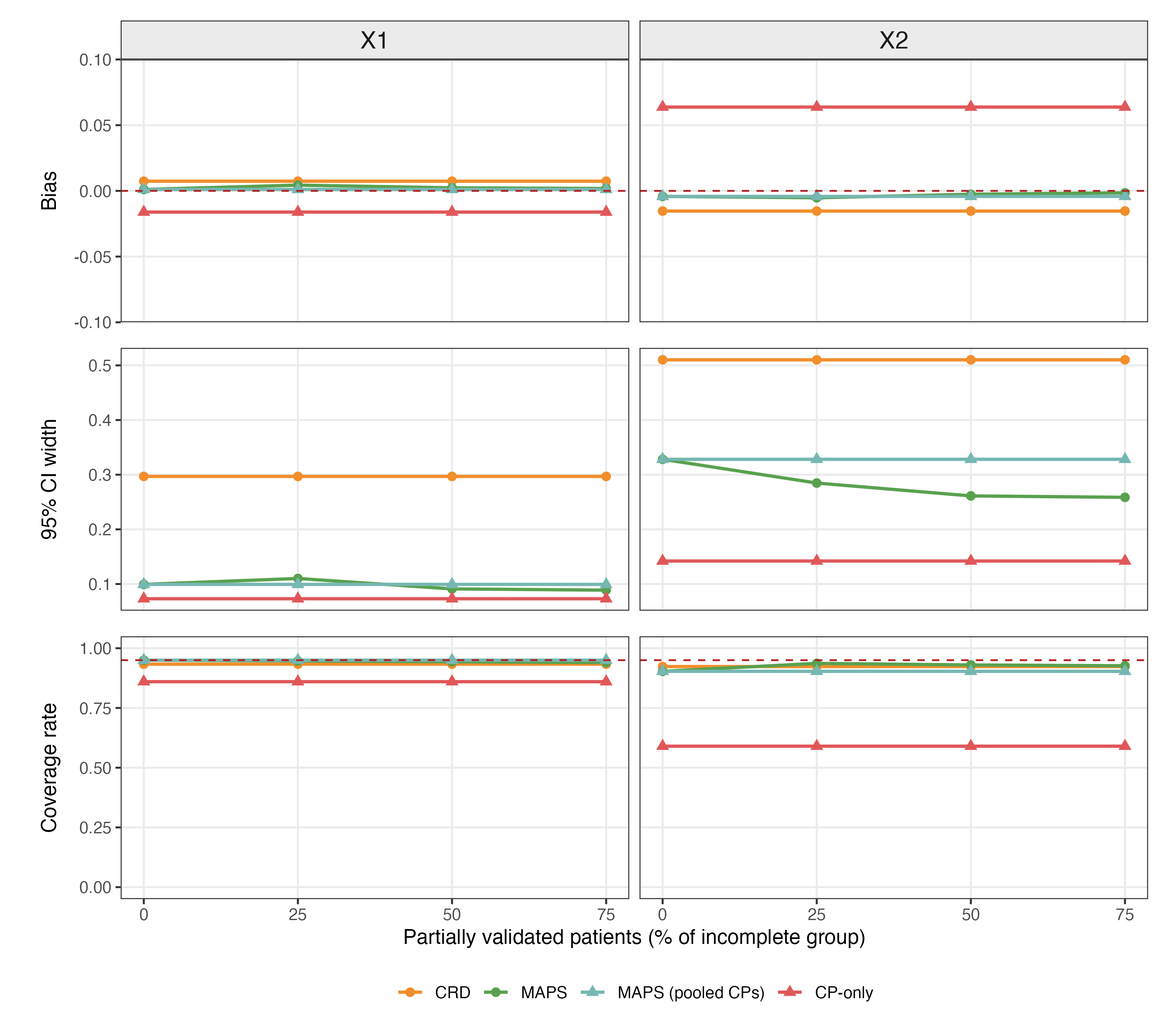}
\caption{Performance across proportions of partially validated patients outside the fully validated subset (0\%, 25\%, 50\%, and 75\%). Each setting used 300 paired replicates with $N=5000$, a target of 500 fully validated patients, outcome CP error $\varepsilon_i\sim N(0,0.05^2)$, and 10\% covariate prediction error. Partial validation was split equally between covariate-only and outcome-only labels. MAPS (pooled CPs) ignored partial labels and used both CP-derived variables outside the fully validated subset. Bias and mean 95\% confidence interval width are on the log-HR scale. Dashed red lines indicate zero bias and nominal 95\% coverage.}
\label{fig:s-partial-label-contribution}
\end{figure}

\clearpage
\paragraph{Contribution of partially validated labels.}
MAPS and CP augmentation that ignored partially validated labels agreed when no partial labels were available (Figure~\ref{fig:s-partial-label-contribution}). With partial validation in 50\% of patients outside $\mathcal D$, MAPS intervals were 8.3\% narrower for $\beta_1$ and 20.4\% narrower for $\beta_2$ than intervals from the comparator. Gains varied by coefficient and allocation: at 25\% partial validation, MAPS intervals for $\beta_1$ were 10.9\% wider, while intervals for $\beta_2$ were 13.2\% narrower. Thus, partial labels improved precision for $\beta_2$ at both allocations, but did not improve precision for every coefficient.

\begin{figure}[!htbp]
\centering
\includegraphics[width=\linewidth]{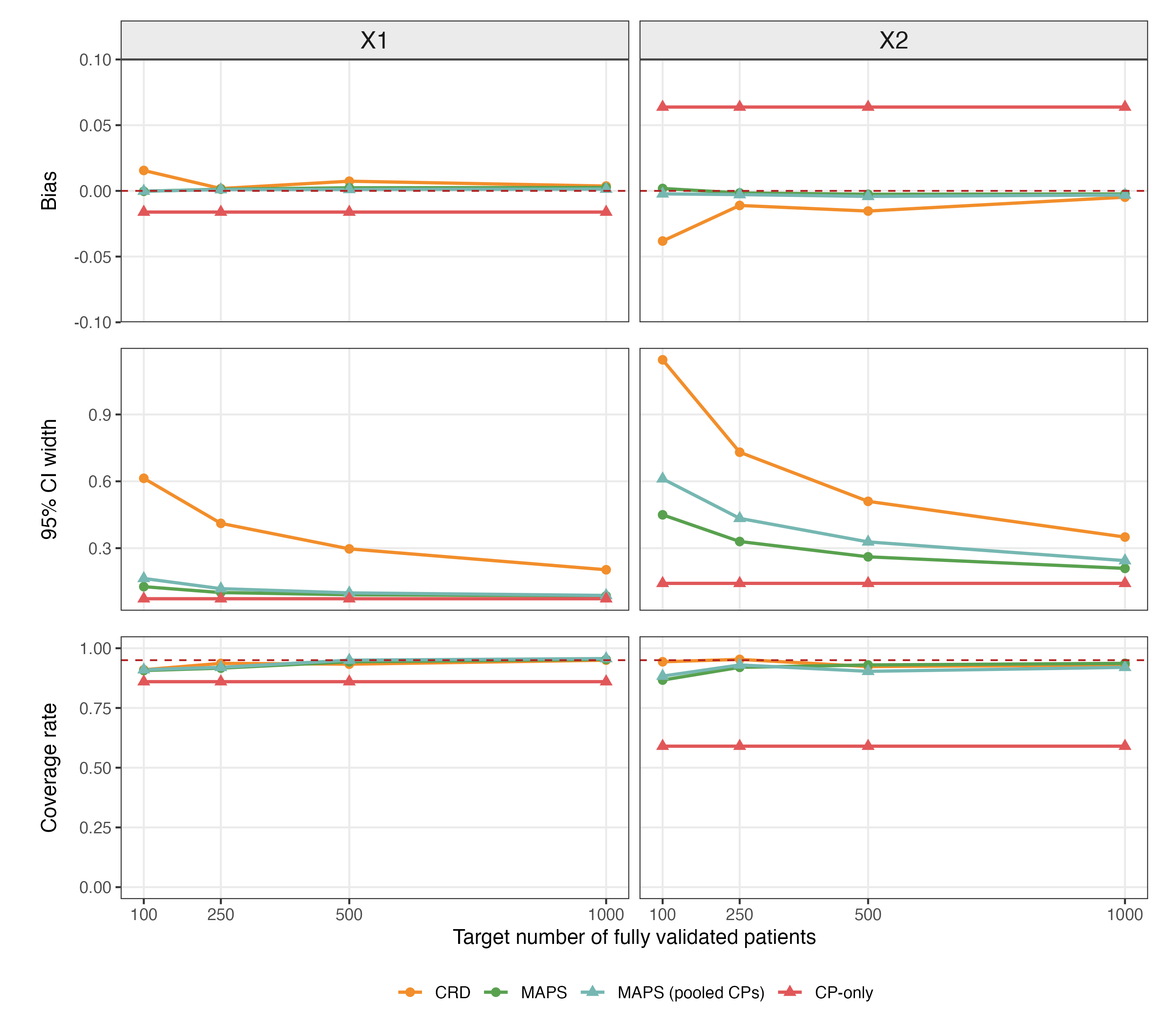}
\caption{Performance across target fully validated sample sizes of 100, 250, 500, and 1000 in cohorts of $N=5000$. Among the remaining patients, 50\% had one variable validated, split equally between covariate-only and outcome-only labels. Each setting used 300 paired replicates, outcome CP error $\varepsilon_i\sim N(0,0.05^2)$, and 10\% covariate prediction error. MAPS (pooled CPs) used both CP-derived variables outside the fully validated subset. Bias and mean 95\% confidence interval width are on the log-HR scale. Dashed red lines indicate zero bias and nominal 95\% coverage.}
\label{fig:s-fully-validated-size}
\end{figure}

\clearpage
\paragraph{Fully validated sample size.}
MAPS retained small average bias as the fully validated subset size varied, but coverage was lower with limited validation (Figure~\ref{fig:s-fully-validated-size}). With approximately 100 fully validated patients, coverage was 90.7\% for $\beta_1$ and 86.7\% for $\beta_2$; the corresponding values were 95.3\% and 93.7\% with approximately 1000 fully validated patients. At the smallest target, the mean estimated standard error for $\beta_2$ was 0.115, compared with an empirical standard deviation of 0.167. The mean effective sample size in $\mathcal D$ was 78.9, and 7.0\% of selected-patient validation probabilities were floored. These results indicate that narrow intervals with small validation subsets should be assessed alongside interval calibration and weight diagnostics.

\clearpage
\bibliographystyle{plainnat}
\bibliography{psppi_survival_jamia_refs}